\documentclass[twocolumn, amsmath,amssymb,english,aps,pra,superscriptaddress]{revtex4-1}
\usepackage[T1]{fontenc}
\usepackage{amsmath}
\usepackage{amssymb}
\usepackage{graphicx}
\usepackage{babel}
\usepackage{color}
\usepackage{braket}
\usepackage[normalem]{ulem}
\usepackage[colorlinks,bookmarks=false,citecolor=magenta,linkcolor=blue,urlcolor=black]{hyperref}
\usepackage{xcolor}

\newcommand{\bea}{\begin{equation}}
\newcommand{\eea}{\end{equation}\noi}
\newcommand{\ber}{\begin{eqnarray}}
\newcommand{\eer}{\end{eqnarray}\noi}
\begin{document}
\begin{center}
\title{Fast charging of quantum battery assisted by noise}
%\\
%or, Noise assisted fast charging of quantum battery}

\author{Srijon Ghosh}
\affiliation{Harish-Chandra Research Institute and HBNI, Chhatnag Road, Jhunsi, Allahabad - 211019, India}

\author{Titas Chanda}
\affiliation{Instytut Fizyki Teoretycznej, Uniwersytet Jagiello\'nski, \L{}ojasiewicza 11, 30-348 Krak\'ow, Poland}

\author{Shiladitya Mal}
\affiliation{Harish-Chandra Research Institute and HBNI, Chhatnag Road, Jhunsi, Allahabad - 211019, India}

\author{Aditi Sen(De)}
\affiliation{Harish-Chandra Research Institute and HBNI, Chhatnag Road, Jhunsi, Allahabad - 211019, India}

\begin{abstract}
We investigate the performance of a quantum battery exposed to local Markovian and non-Markovian dephasing noises. The battery is initially prepared as the ground state of a one-dimensional transverse $XY$ model with open boundary condition and is charged (discharged) via interactions with local bosonic reservoirs. We show that in the transient regime, quantum battery (QB) can store energy faster and has a higher maximum extractable work, quantified via ergotropy,  when it is affected by local phase-flip or bit-flip \emph{Markovian} noise compared to the case when there is no noise in the system.  In both the charging and discharging processes, we report the enhancement in work-output as well as in ergotropy when all the spins are affected by \emph{non-Markovian} Ohmic bath both in the transient and the steady-state regimes, thereby showing a counter-intuitive advantage of decoherence in QB. Both in Markovian and non-Markovian cases, we identify the system parameters and the corresponding noise models which lead to maximum enhancement of work-output and ergotropy. Moreover,  we show that the benefit due to noise persists even with the  initial state  being prepared at a moderate temperature. 
    
\end{abstract}
\maketitle
\end{center}
\section{Introduction}

%With the progress of civilization, modern gadgets like medical aids,  cooking and entertainment devices, computers, mobiles  play a crucial role in fulfilling our daily necessities. The smooth running of most of these appliances require  an energy storage device, e.g.  batteries which are typically  of the electrochemical type, converting chemical energy to the electrical one.
% Due to the limitations in the availability of conventional resources, it is of prime importance to improvise a novel kind of resource, keeping in mind the constant efforts to scaled-down the appliances. Both the purposes 
In the last century,  quantum mechanical description of nature was shown to improve the performance of  the devices compared to the analogous  classical devices and at the same time, fulfilled the effort to miniaturize them  \cite{NielsenChuangbook}. It has also been established that  quantum properties,  like coherence \cite{pleniorev}, entanglement \cite{horodecki, amader_ent}, quantum discord \cite{modiamader} are responsible  to enhance the capability in communication \cite{revcrypto}, computation \cite{oneway}, optimal control \cite{optcontrol}  etc. Moreover, many of these discoveries have  successfully been realized in laboratories \cite{photonRMP, ionRMP}.

Towards this aim, recently
%Towards this aim, it is natural to build 
a quantum-version of an energy-storage device, a battery, was proposed \cite{alicki, campaioli'18}.  Such a consideration is also a part of  an active field of research, called quantum thermodynamics \cite{qthermobook, horojona}, which for example, deals with  whether thermodynamical principles maintain their validity or get modified in the quantum mechanical domain or not.     
The idea of quantum batteries has been introduced with the aim of obtaining advantages in the process of charging and discharging. It was shown  \cite{alicki} that the entangling operations are required to enhance extractable work stored in a quantum battery (QB) which consists of an arbitrary number of  independent and identical quantum systems. Interestingly, it was observed that the required entangling operations do not necessarily generate entanglement in the system to maximize the storage of energy  \cite{acin, binder, andolina1}. Moreover,   there can be a model of QB from which  a high amount of power-extraction is possible without generating entanglement in the system \cite{modisai'17}. 
%It was later shown that to obtain increased power generation of discord is necessary \cite{}.
Although  resources responsible for showing quantum advantages in QB  is  still not properly understood \cite{Qvscl} --
%it is clear that the battery made on  the principles of quantum theory can accomplish the need of current development in technologies and hence  
characterizing it via the various figures of merit is essential. 

Towards implementing quantum batteries in realizable systems,  solid-state QB employing Dicke state has been prescribed \cite{ferraro}, where extensive quantum advantage has been obtained due to the  interaction between  the system  and the common reservoir, induced by collective charging compared to parallel charging of the  qubits. In a similar spirit,  ordered quantum spin chain which can be charged via a local magnetic field has been shown to be a potential candidate for QB \cite{lemodi'18, andolina'18, rosa19} (see also \cite{Crescente}).  Recently, some of us have shown that  disordered interactions, instead of having detrimental effects,  can enhance the  extraction of power from the QB \cite{srijon'20} (see also \cite{rossini'19}).

During the charging-discharging process of QB in laboratories, one of the possible obstacles is the decoherence  \cite{opensysbook} which can, in general,  decrease the performance of the device.  In very recent times, studying the dynamics of QB-system in presence of an environment \cite{farina'19, alickiopen, latune'19, kamin'19, zakavati'20, Gherardini'20, munro'20, sibai_2020, mitchison, carrega}  has created lots of attention and has revealed several interesting  characteristics including the effect on ergotropy due to the interplay between a coherent and an incoherent source of the charger, i.e.,  between the unitary and the non-unitary processes,  
%\tc{(TO BE MODIFIED): Why? don't understand}, 
the stability of the battery, the superextensive capacity of the battery by using  dark states \cite{munro'20}, and enhanced charging and preserving of energy via non-Markovian dynamics   \cite{kamin'19}. Note here that states evolved in the presence of  non-Markovian noisy environment was shown to be useful for generating (or enhancing) certain quantum features, such as the  periodic revival of quantum entanglement,  quantum coherence etc., necessary  for building quantum technologies \cite{breuer'09, laine'10, rivas'10, lu'10, haikka'13, Chin'12, vasile'11, thorwart'09, huelga'12, schmidt'11, sabrinarev, titassamya, rivu'20}.  

\begin{figure}
\includegraphics[width=\linewidth]{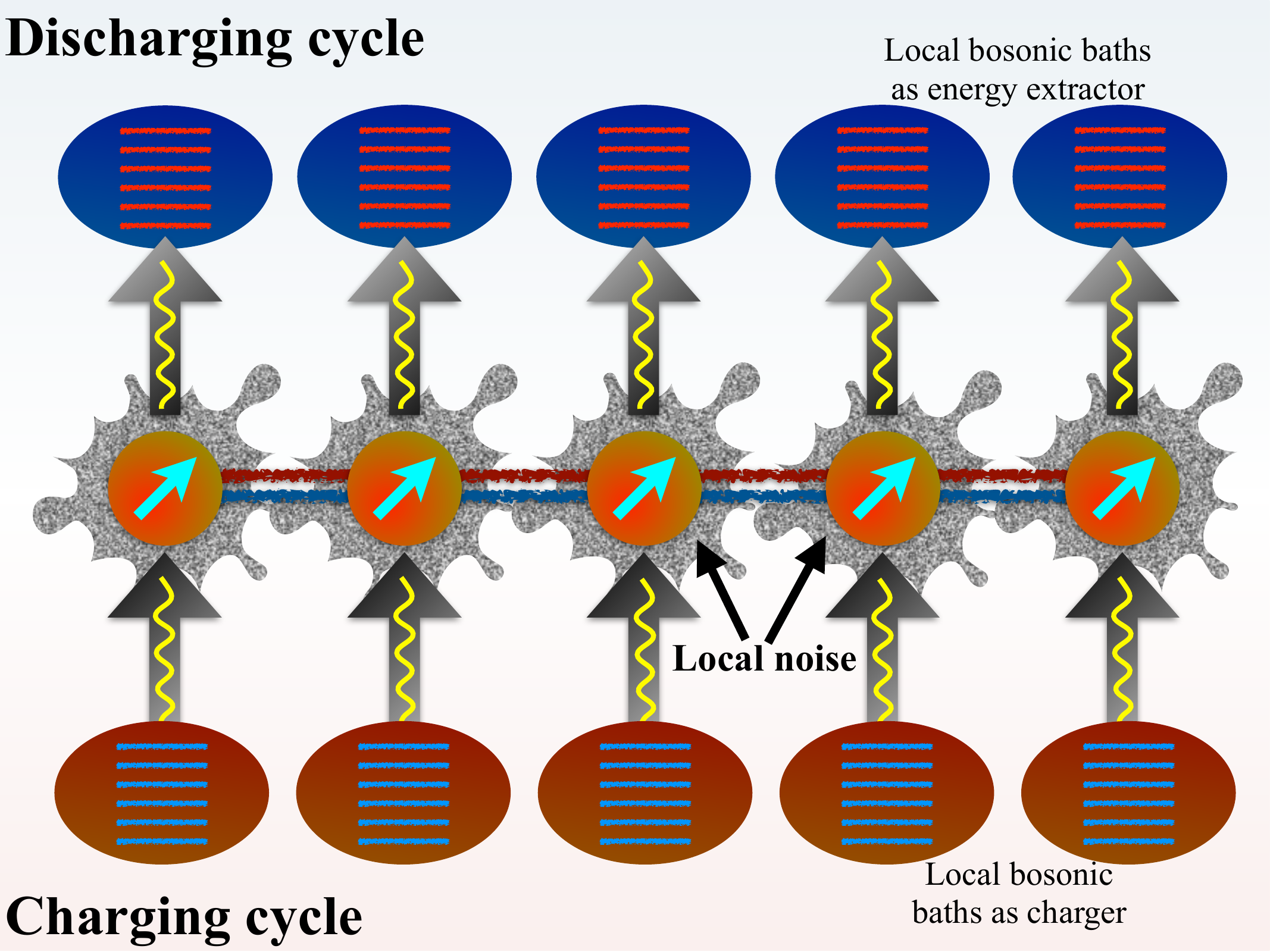} 
\caption{(Color online.) \emph{Schematic  model of the quantum battery.} An  interacting quantum spin model with open boundary condition acts as a battery, while  charging and discharging processes  take place by using local bosonic reservoir. Moreover,  the local noise (Markovian as well as non-Markovian)  acts on each spins (grey clouds).}
\label{fig_schematic}
\end{figure}

In the present work, we consider a one-dimensional $XY$ spin model with a transverse magnetic field  in open boundary condition as a quantum battery. Every  individual spin of it is connected to two sets of local bosonic reservoirs -- one acts as a charger while the other one  is an energy extractor responsible for   the discharging process. In addition, we  consider dephasing noise which acts locally on some of the spins or on all the spins of the model (see  Fig. \ref{fig_schematic} for the schematic of the model). In the Markovian domain, we find that during the storing of energy and in terms of maximal gained  work, i.e., the ergotropy, there exists a crossover time upto which noise helps to  pump more energy (work-gained) in QB compared to that of the noiseless case, while non-Markovianity further enhances the capacity of the charging process in the entire time period. In the later case, the work-gained  as well as the amount of extractable work by the QB  also increase when more number of spins are exposed to noise.
%{increase of the number of spins on which noise acts}. \textcolor{blue}{TC: mane ta ektu annorokom lagchhe, `increase of the number of spins' focus pachche.}  
 
We analytically show the advantage obtained for Markovian bit-flip and phase-flip noises when the system is prepared in the ground state of the transverse Ising spin chain consisting of two spin-1/2 systems. We also find the range  of parameters for which bit-flip noise leads to a higher amount of work than that of the phase-flip ones.
 We then illustrate that such a hierarchy remains valid also for the ergotropy in the intermediate time of the dynamics.   We observe that the  bit-flip noise performs better in terms of  the work-output if the initial state is chosen from the paramagnetic phase compared to that of the antiferromagnetic one, both in the presence of Markovian and non-Markovian noises while we demonstrate that  such a phase dependence does not appear in the computation of ergotropy to show the advantage of the bit-flip noise over the phase-flip ones. In both the noisy scenarios, the enhancement per spin is independent of the system size. During discharging,   non-Markovian noise turns out to  be more efficient  than that of the Markovian dynamics as well as in the absence of noise. Moreover, we notice that the advantage  due to noise can be obtained in terms of work-output, as well as ergotropy,
both for the  ground and  the canonical equilibrium states  of the transverse Ising model having moderate temperatures as the initial states.
% is either in the 

 The paper is organized as follows. In Sec. \ref{sec_formalism},  we discuss the formalism of quantum battery and its dynamics considered here.   The consequence of Markovian noise on QB based on transverse Ising model consisting of two spins is studied analytically in Subsec. \ref{subsec_QBmarko2}, while  we present the results about  the effects of other parameters and  Markovian noise on work and ergotropy of  QB having more number of spins in Subsec. \ref{subsec_QB_marko4}. Sec. \ref{sec_QB_nonMarko4} deals with non-Markovian noise and its advantage in the work and ergotropy generation as well as their  extraction in QB.  Finally, we end with concluding remarks in Sec. \ref{sec_conclu}.

\section{Quantum battery  exposed to noisy environment: Preliminaries }
\label{sec_formalism}

A quantum battery is a collection of a \(N\)-body system, used to store energy, which can be extracted in a suitable later time. Sepcifically, the ground  or the canonical equilibrium  state of  an interacting Hamiltonian, \(H_B\),  can be chosen as the initial state ($\rho_{0}$) of the battery. If we assume the QB and the charging device, which can be a local magnetic field, are isolated from the environment,  the charging-discharging process of QB can be well-described by the unitary dynamics \cite{lemodi'18, andolina'18, srijon'20}, and after a certain time, \(t\), total work-gained by the QB can de defined as 
\begin{equation}
\label{eq_work}
W(t) = \mbox{Tr}(H_B \rho_{t})-\mbox{Tr}(H_B\rho_{0}),
\end{equation}  
where \(H_B\) is the Hamiltonian of the QB, and $\rho_{t} = U\rho_0 U^{\dagger}$ is the time-evolved state of the battery with \(\rho_0\) being the initial state. Note that  extracting energy from the QB in this scenario is the reverse process of the charging one which will not be the case for open quantum dynamics.

In this paper, we mainly focus on the effect of the environment on the generation (and extraction) of energy in the QB. The Hamiltonian of an interacting quantum spin model considered as a battery can be described as
\begin{equation}
\label{eq_batterymodel}
H_{B}=\frac{h}{2}\sum_{j=1}^N  \sigma_{j}^z+\frac{J}{4}\sum_{j=1}^N [(1+\gamma)\sigma_{j}^x\sigma_{j+1}^x + (1-\gamma)\sigma_{j}^y\sigma_{j+1}^y] ,
\end{equation} 
where \(\sigma^i\)s, \( i=x, y, z\) are the Pauli spin matrices, \(N\) is the total number of spins,  \(h\) and \(J\) respectively represent the strength of the magnetic field and nearest-neighbor coupling constant, and \(\gamma \geq 0\) is the anisotropy parameter \cite{Sachdevbook}. In the thermodynamic limit,  the model undergoes a quantum phase transition  at \(\lambda \equiv J/h =1\) -- it is in a paramagnetic phase when \(|\lambda| \leq 1\) while in an antiferromagnetic phase with \(\lambda >1\) and a ferromagnetic phase with \(\lambda <-1\). 

During the charging process, the QB is attached to a set of local bosonic reservoirs, which acts as an absorption channel  (an effective dynamical map after integrating-out the bosonic degrees of freedom) that is responsible to supply energy to the system, as depicted in Fig. \ref{fig_schematic}. After the completion of the charging process, i.e., when the energy in QB does not change with time,
the discharging process starts, when the spins of the QB are attached to a different set of local bosonic reservoirs, which extracts the energy by acting as a
dissipation channel (which can again be an effective dynamical map for the spins after integrating-out the bosonic degrees of freedom). Moreover, during these two processes, each spin of QB can be exposed to dephasing noise -- (a) phase-flip and (b) bit-flip noise. Therefore, the reduced dynamics of QB can be described by Lindblad-Gorini-Kossakowski-Sudarshan \cite{lindblad'76, gorini'76, opensysbook} master equation as
\begin{eqnarray}
\label{eq_master}
\frac{d\rho(t)}{dt} &=& -i [H_B,\rho(t)] \nonumber\\
 &+& \sum_{k}\gamma_{(k)} (L_{k,j}\rho(t)L_{k,j}^{\dagger}-\frac{1}{2}\lbrace L_{k,j}^{\dagger}L_{k,j},\rho(t) \rbrace), 
\end{eqnarray}
where the dimension of \(\gamma_{(k)}\)s are chosen to be in the units of inverse of time to make Lindblad operators, \(L_{k,j}\)s, dimensionless. There are three kinds of \(L_{k,j}\) applied at the \(j\)-th spin,  given by
\begin{eqnarray}
L_{1,j} =  \sigma_j^{+},\,  L_{2,j} =  \sigma_j^{-}, \, L_{3, j} =    \sigma_j^i, \, \, i=z, x, 
\end{eqnarray}
for absorption (during charging process), dissipation (during discharging process) and dephasing channels respectively,
and the corresponding coefficients read as
\begin{eqnarray}
\gamma_{(1)} =  \Gamma_{abs}, \, \gamma_{(2)} =  \Gamma_{dis},\,\,  \gamma_{(3)} = \Gamma_{dph}^i \, \, i=z, x,
\label{eq_lindbladop}
\end{eqnarray}
where $\sigma^{+} = \frac{\sigma_{x}+i\sigma_{y}}{2}$,   $\sigma^{-} = \frac{\sigma_{x}-i\sigma_{y}}{2}$ and \(\Gamma_{abs}\), \(\Gamma_{dis}\) and  \(\Gamma_{dph}^i,\, i = x, z \) denote respectively the rate of absorption when QB consumes energy from the charger, the rate of dissipation during discharging, and the strength of the dephasing noise. 
%Note, that when the discharging from the QB takes place, the hermitian conjugate of \(L_{k,j}\) has to be considered and in that case, \(\Gamma_{abs}\) is replaced by \(\Gamma_{dis}\). 
    
% , respectively  given by
%\begin{eqnarray}
%\label{eq_noise}
%H^{1}_{noise} = \Gamma^{z}_{dph} \sum_{j}\sigma_{j}^{z}, \,\, H^{2}_{noise} = \Gamma^{x}_{dph} \sum_{j}\sigma^{x}_{j},
%\end{eqnarray}  
%where \(\Gamma_{dph}^i, \,\, i =z, x \) is the strength of the noise.  
We consider two scenarios -- \\
Case 1. \emph{ Time-independent noise.} Each spin of QB interacts with the bosonic reservoirs, and at the same time, with local dephasing noise, either in the $x$ or in the $z$-direction. All the \(\gamma_{(k)}\)s are time independent. The dynamics of QB in this case is Markovian.  \\
Case 2. \emph{Time-dependent noise.}  \(\gamma_{(k)},\, \, k =1, 2\) are considered to be time independent but the strength of the dephasing noise, i.e., \(\gamma_{(3)}\) is taken as a time-dependent variable, leading to non-Markovian effects in dynamics which will be discussed in details in the succeeding section.

%Fixing a proper initial state of the QB,   the time-evolved battery state, $\rho_{t}$ at a given instance of time  can be obtained by solving the master equation, given in Eq. (\ref{eq_master}),  and hence the total work-output  of the QB affected by noisy environment can also be computed by using  Eq. (\ref{eq_work}). To quantify the rate of change of work, we  also investigate the instantaneous power,   defined as
%\begin{equation}
%\mathcal{P}_{ins} = \lim_{\bigtriangleup t \rightarrow 0} \frac{|W(t+\bigtriangleup t) - W(t)|}{\bigtriangleup t} = |\frac{dW}{dt}|. 
%\end{equation}
To quantify the maximal amount of work that one can gain (extract) from the battery at any time instant $t$, we calculate ergotropy, denoted by $Erg(t)$,  which is defined as
\begin{equation}
\label{ergotropy}
Erg (t) = E_{B}(t) - \min_{U_{B}} \text{Tr}[H_{B} U_{B} \rho(t) U_{B}^\dagger].
\end{equation}
Here $\rho(t)$ and $E_{B}(t)$ is the time evolved state of the system and energy of the battery at time instant $t$ respectively. $U_{B}$'s are the unitaries over which minimization occurs to get the extractable work from the system \cite{alicki}. It can be computed by using spectral decompositions of \(\rho(t)\) and \(H_B\) \cite{farina'19}.

\section{Noise-induced enhancement  in a  quantum battery  based on a transverse Ising chain without memory}
\label{sec_QBMarko}

We will now show that Markovian noise can help to give certain improvement in the charging-discharging duo,  when the battery is in the ground state or the thermal state of the transverse Ising model. In Subsec. \ref{subsec_QBmarko2}, we will  analytically show that  QB consists of two spins, both phase-flip and bit-flip noise acting on QB can give increment in total work-output and ergotropy  during the initial periods of the charging process compared to the scenario where the noise is absent. We will then report the improvement of noisy QB composed of  the transverse Ising model consisting of higher number of spins even in presence of Markovian noise  in Subsec. \ref{subsec_QB_marko4}.  

\subsection{ Noise-assisted battery constructed via  transverse Ising chain of two spins: Markovian regime}
\label{subsec_QBmarko2}

Let us suppose that  quantum battery consists of  two spin-1/2 particles, described by the Hamiltonian in Eq. (\ref{eq_batterymodel}) with unit anisotropy, \(\gamma =1\), i.e., the  transverse Ising model. The battery is initially prepared as the ground state of the  model. Let us first concentrate on the QB when  phase-flip channel acts on each spin of the battery independently. In this case, we obtain the  following result:\\

\noindent\textbf{Proposition 1.}
\label{th_phflip}
%\textbf{Theorem.} 
In the transient regime, the work-gained by the  transverse Ising model-based QB with the help of local bosonic reservoirs is higher in presence of  local phase-flip noise than the one without the influence of noise. \\ 
%\end{theorem}
\textbf{Proof.}
The master equation in Eq. (\ref{eq_master}) for the noise in the \(z\)-direction can explicitly be written as
\small
\begin{eqnarray}
& \frac{d\rho}{dt} & \ = -i[H_B,\rho(t)] \nonumber \\
&+& \Gamma_{abs}[(\sigma^{+}\otimes I) \rho(t) (\sigma^{-}\otimes I) - \frac{1}{2} \lbrace (\sigma^{-}\otimes I) (\sigma^{+}\otimes I) , \rho(t) \rbrace \nonumber \\
&+& (I \otimes \sigma^{+}) \rho(t) (I \otimes \sigma^{-}) - \frac{1}{2} \lbrace ( I\otimes \sigma^{-}) ( I\otimes\sigma^{+}) , \rho(t)\rbrace] \nonumber \\
&+& \Gamma_{dph}^z [(\sigma_{z} \otimes I) \rho(t)(\sigma_{z} \otimes I) + (I \otimes \sigma_{z}) \rho(t) (I \otimes \sigma_{z})  - 2\rho(t) ], \nonumber \\
\label{eq_master2q}
\end{eqnarray}
\normalsize
where the charger and the noise act locally. Let us consider the density matrix of the initial state, $\rho_0$, of the battery as the ground state of the transverse Ising  Hamiltonian $H_B$, given by 
\begin{equation}
\label{eq_gr}
\rho_0= \begin{pmatrix}
\frac{p^{2}}{(1+p^{2})} & 0 & 0 & \frac{p}{(1+p^{2})}\\
0 & 0 & 0 & 0\\
0 & 0 & 0 & 0\\
\frac{p}{(1+p^{2})} & 0 & 0 & \frac{1}{(1+p^{2})}\\
\end{pmatrix}, 
\end{equation}
where $p= \frac{4(h+e_0)}{J}$ and $e_0 = - \sqrt{h^2 + \frac{5 J^2}{8}}$ is the ground state energy  of $H_B$. Solving Eq. (\ref{eq_master2q}) (see Appendix \ref{sec_appendix}), we can obtain the time-evolved  state of the  battery, \(\rho_t\). Since the two-body Ising Hamiltonian of the QB  in the computational basis takes the form of a ``X''-matrix, given by 
\begin{equation}
\label{eq_HBsimple}
H_B = \begin{pmatrix}
h & 0 & 0 & \frac{J}{2}\\
0 & 0 & \frac{J}{2} & 0\\
0 & \frac{J}{2} & 0 & 0\\
\frac{J}{2} & 0 & 0 & -h\\
\end{pmatrix},
\end{equation}
 a compact form of extracted work in presence of phase-flip noise can be computed analytically. Since we want to calculate the effects of noise on the dynamics of \(W(t)\), we  denote the quantity as  $W(\Gamma_{dph}^z)$, omitting its functional dependence on \(t\) from notation. The straightforward calculation leads us to
 \begin{eqnarray}
 \label{eq_nyakadeph}
 && W (\Gamma_{dph}^z) = \frac{e^{- t\Gamma_{abs}}}{4\Gamma_{dph}^zA} \times\nonumber\\
  && [4Jp B  \Gamma_{dph}^z + h \lbrace 2J p(B-1)+ 4\Gamma_{dph}^z(e^{ t\Gamma_{abs}}A -2) \rbrace ],\nonumber \\
\end{eqnarray}  
with \(A= (1+p^2)\) and \(B = e^{-4t \Gamma_{dph}^z}\).
On the other hand, in the noiseless case, i.e., for  $\Gamma_{dph}^z =0$, the extracted work, $W(\Gamma_{dph}^z=0)$, is given by
\begin{equation}
\label{eq_nonyakadeph}
W(\Gamma_{dph}^z=0) = h + \frac{e^{-t \Gamma_{abs}}[2Jp-2h(2+Jpt)]}{2A}.
\end{equation}
By subtracting Eq. (\ref{eq_nonyakadeph}) from  Eq.  (\ref{eq_nyakadeph}), and  expanding them around $t=0$, we get
\begin{equation}
\delta_{\mbox{ph-f}}^{adv} \equiv W (\Gamma_{dph}^z)-W(0) = - \frac{8 \Gamma_{dph}^z (1+e_0) t}{(1+p^2)} + \mathcal{O}(t^2).
\end{equation}
Keeping  the first-order term in \(t\) (and ignoring the higher order  ones), we find that  \(\delta_{\mbox{ph-f}}^{adv} \)  is a positive number (since $e_0$ is a negative number and its absolute value is greater than $J$), thereby implying the advantage in a noisy scenario than the noiseless one in  a transient regime.  
%\end{proof}
$\blacksquare$

The above Proposition shows that after the evolution starts, noise in the $z$-direction helps to store more energy in the QB of the transverse Ising model than that in the noiseless situation. Instead of phase-flip noise, if both the spins are affected by the bit-flip noise, the noisy situation still remains advantageous with the restriction in the parameter-space of the transverse  Ising chain as shown in the Proposition below. \\
\textbf{Proposition 2.}
\label{th_bitflip}
When local bit-flip noise acts on both the spins of QB which is initially in the ground state of a transverse Ising chain,  the work-output, in the transient regime, gives higher value in the noisy case than that of the noiseless one, provided  the initial state is prepared with $\lambda \lesssim 0.89$.  \\
% \end{theorem}
%Putting noise in the $x$ direction GSKL master equation takes the form,
%By putting noise in the $x$ direction, the extractable work $W^{x}(\Gamma_{deph})$ take the form,
{\bf Proof.} For  bit-flip noise, we can rewrite the  master equation as in Eq. (\ref{eq_master2q})  by replacing \(\sigma^z\)  by \(\sigma^x\) and the strength of the noise by \(\Gamma_{dph}^x\) (see Appendix \ref{sec_appendix} for details).  If we take the ground state as the initial state of \(H_B\), given in Eq. (\ref{eq_gr}),  and compute the work-output during charging process, we obtain

\begin{eqnarray}
\label{eq_bitflipW}
&& W(\Gamma_{dph}^x) = \frac{1}{4\Gamma_{dph}^x A (\Gamma_{abs} +2\Gamma_{dph}^x)} \times \nonumber \\
&& e^{-t (\Gamma_{abs} + 4 \Gamma_{dph}^x)} [A'-B' e^{2t \Gamma_{dph}^x}+e^{4t\Gamma_{dph}^x}( C'e^{t\Gamma_{abs}}+ D')],\nonumber\\
\end{eqnarray} 
where
$A' = 2Jhp(\Gamma_{abs} +2 \Gamma_{dph}^x)$, $B' = 8h\Gamma_{dph}^x(\Gamma_{abs} +\Gamma_{dph}^x-p^2 \Gamma_{dph}^x )$, $C' = 4\Gamma_{abs} h\Gamma_{dph}^x(1+p^{2})  $ and $D' = Jp(\Gamma_{abs} +2\Gamma_{dph}^x)(2\Gamma_{dph}^x-h) $ while the \(W (\Gamma_{dph}^x =0)\) is same as in Eq. (\ref{eq_nonyakadeph}). 
In the transient regime,  expansion of the  exponential  at \(t=0\) gives 
\begin{equation}
\delta_{\mbox{bit-f}}^{adv} \equiv W(\Gamma_{dph}^x) - W(0) = 2\Gamma_{dph}^x t \frac{(1-p^{2})}{1+p^{2}} + \mathcal{O}(t^2).
\label{eq_bfadv}
\end{equation}
As before, we only keep the above series up to the first order in \(t\). To get the  improvement in  energy stored in presence of noise along the $x$-direction, $p^{2}$ should be less than $1$ which holds when  $0 < \lambda \lesssim 0.89$.   %$x$ direction of applied dephasing channel is more profitable (in the sense of work extraction from the bettery) than when no interaction with environment is present.
Unlike phase-flip noise, we find here that in the transient regime, the work can be enhanced in presence of noise  when the initial state is the ground state of the transverse Ising model with suitably chosen parameters. $\blacksquare$

The above two Propositions also lead to a comparison between the work-outputs   when the amount of  noise  applied to  the $x$- and the $z$-direction respectively. Suppose the amount of noise in both the cases  are same, i.e 
\(\Gamma_{dph}^z = \Gamma_{dph}^x = \Gamma_{dph} \). \\
\textbf{Proposition 3.}
The amount of  stored energy in the transient region is more in case of bit-flip noise than that of the scenario with the phase-flip noise, provided the initial state is prepared  from the paramagnetic phase of the transverse Ising chain, far away from the critical point. \\
\noindent\textbf{Proof.} From Eqs. (\ref{eq_nyakadeph}) and (\ref{eq_bitflipW}), we immediately find (up to the first-order approximation of \(t\))
\begin{equation}
\delta^{adv}_{x>z} \equiv W(\Gamma_{dph}^x) - W(\Gamma_{dph}^z) = 2 \Gamma_{dph} t \frac{(1+\lambda p-p^{2})}{1+p^{2}},
\label{eq_comparison}
\end{equation}
which is positive only when  $(1+\lambda p) > p^{2}$. It implies that  more amount of  work is generated in the bit-flip noise   than that of the phase-flip ones if  $\lambda \lesssim 0.62$. When $\lambda \gtrsim 0.62$, noise in the $z$ direction turns out to be more beneficial than that of the bit-flip one as we will also see from the numerical simulations with more number of spins in the next section. 
$\blacksquare$

\begin{figure}
\includegraphics[width=\linewidth]{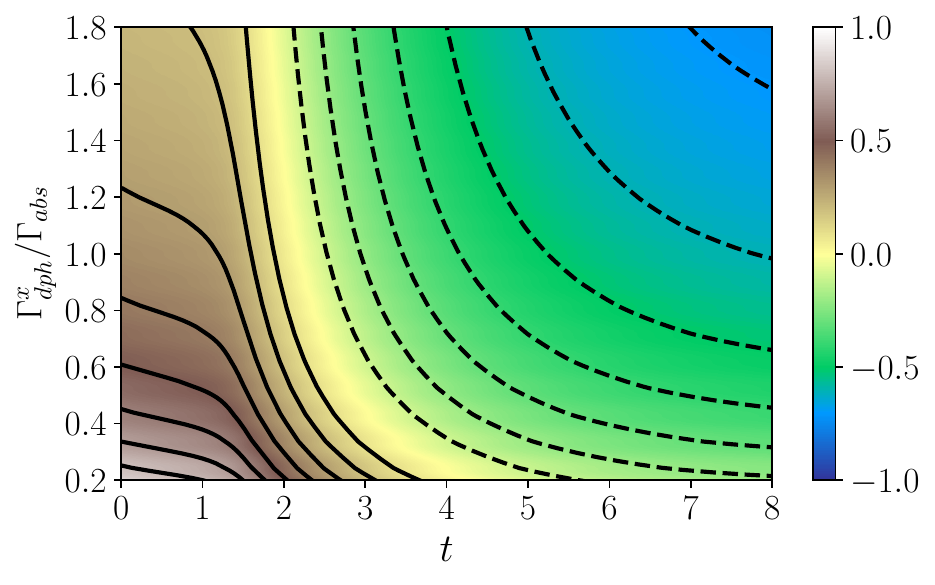} 
\caption{(Color online.) \emph{Role of noise-absorption ratio.}
Contour of the advantage in ergotropy, \(\Delta^{erg-adv}_{bit-f} = Erg(\Gamma_{dph}^x)-Erg(0)\),  with respect to time, \(t\),  (abscissa) and \(\Gamma_{dph}^x/\Gamma_{abs} \) (ordinate). Different contours represent the fixed value of   \(\Delta^{erg-adv}_{\mbox{bit-f}}\). Here \(N=4\) and the initial state is  in the ground state of the transverse Ising model with \(\lambda \equiv J/h = 0.5\). Both the axes are dimensionless.}
\label{fig_advantage_ratio}
\end{figure}

Let us now perform the similar analysis for ergotropy and find out whether noise really helps to produce more ergotropy, when the initial state of the battery is the ground state of the transverse Ising model. In this case, we notice that the ergotropy shows noise-induced enhancement in the intermediate time-interval, and so the expansion at \(t=0\) performed for work-output is not valid in this situation, thereby making the analytical treatment difficult. However, we find that for the noiseless situation, \(Erg(t=1.5) = 0.398\), while when dephasing noise along the $x$-direction acts on the first spin as well as on both the spins, the numerical values of ergotropy turn out to be $0.417$ and $0.435$ respectively, showing clear increment over that of the noiseless case for $\lambda=0.5$. However, when we apply phase-flip noise on a single spin and both the spins, the values of ergotropy become $0.403$ and $0.407$ respectively, thereby showing that the effects of  bit-flip noise on the performance is more pronounced  than that of the phase-flip ones in certain parameter values and time-window. Moreover,  Fig. \ref{fig_advantage_ratio} depicts such advantage in ergotropy in case of bit-flip noise for the initial period of the dynamics, when the the battery is initially prepared as the ground state of the transverse Ising model (see the next subsection for more details).
%Hence, although both kind of noises are actually giving advantage over noiseless scenario, but bit-flip is better than phase-flip one.

%\textcolor{blue}{can u give ergotropy value for the above for bit flip noise. What is the lambda value?}

\subsection{Battery made up of  transverse Ising model with arbitrary spins: Markovian dynamics}
\label{subsec_QB_marko4}
Let us now move to the QB, built up with \(N (>2)\) number of interacting spin-1/2 particles, governed by the transverse Ising  Hamiltonian, \(H_B\) with \(\gamma =1\). 
For a  better comparison between different scenarios, we normalize the spectrum of  the Hamiltonian in the rest of the paper as
\begin{eqnarray}
(H_B - e_{0} I) / (e_{\max} - e_{0}) \rightarrow H_B,
\label{eq:normalizing}
\end{eqnarray}
where $e_{0}$ and $e_{\max}$ are the lowest and highest energy levels of the former Hamiltonian. Due to this normalization, $W$ is restricted between $0$ and $1$, fixing the unit of energy as well as of time, irrespective of the system parameters.

In this subsection, our aim is to confirm  whether the results obtained in the previous subsection with two spins also hold when the number of spins in the QB designed via transverse Ising model  increases. To find the optimal work-gained or ergotropy of the battery, we have to set all the parameters involved in the charging-discharging process in a \emph{proper} manner. From the set-up of the problem, we find that  \(W\) (\(Erg\)) is  a function of \(\lambda, \gamma, \Gamma_{abs}, \Gamma_{dph}^i,\, (i=x, z)\), apart from its dependence on time.  To choose parameters  suitably,  we now  carefully analyze the dependence of \(Erg\) on the parameter space. Qualitatively, similar dependence can also be found for \(W\). 

\subsubsection{\textbf{Noise-absorption (dissipation) ratio.}} Let us first identify the role of \(\Gamma_{dph}^x/\Gamma_{abs}\) in the charging process. By fixing the initial state as the ground state of the transverse Ising chain and by choosing \(\lambda =0.5\) from the paramagnetic phase, we investigate the enhancement of ergotropy  due to the introduction of bit-flip noise, \(\Delta^{erg-adv}_{bit-f} = Erg(\Gamma_{dph}^x)-Erg(0)\), with the variation of  time and \(\Gamma_{dph}^x/\Gamma_{abs}\), as depicted in Fig. \ref{fig_advantage_ratio}. The number of spins in the Ising chain is taken to be four and all the spins are affected by local bit-flip noise. As shown in Proposition 2 for the amount of stored energy in the QB at the initial period of time, ergotropy mimics the similar behavior. We also observe that the amount of increment 
remains sufficiently high for sufficiently long time
 when \(0.2\lesssim \Gamma_{dph}^x/\Gamma_{abs} \lesssim0.4\) which can be referred to as the optimal operating region for the QB.  When the noise is applied in the \(z\)-direction, we also notice that  with \(0.2\lesssim \Gamma_{dph}^z/\Gamma_{abs} \lesssim0.4\),  \(\Delta^{erg-adv}_{ph-f} >0\) although   unlike the noise in the \(x\)-direction, the increment is not prominent in this case.  Motivated by these observations, we have chosen the ratio to be \(0.3\), for both the noisy cases,  throughout the paper.

\begin{figure}
\includegraphics[width=\linewidth]{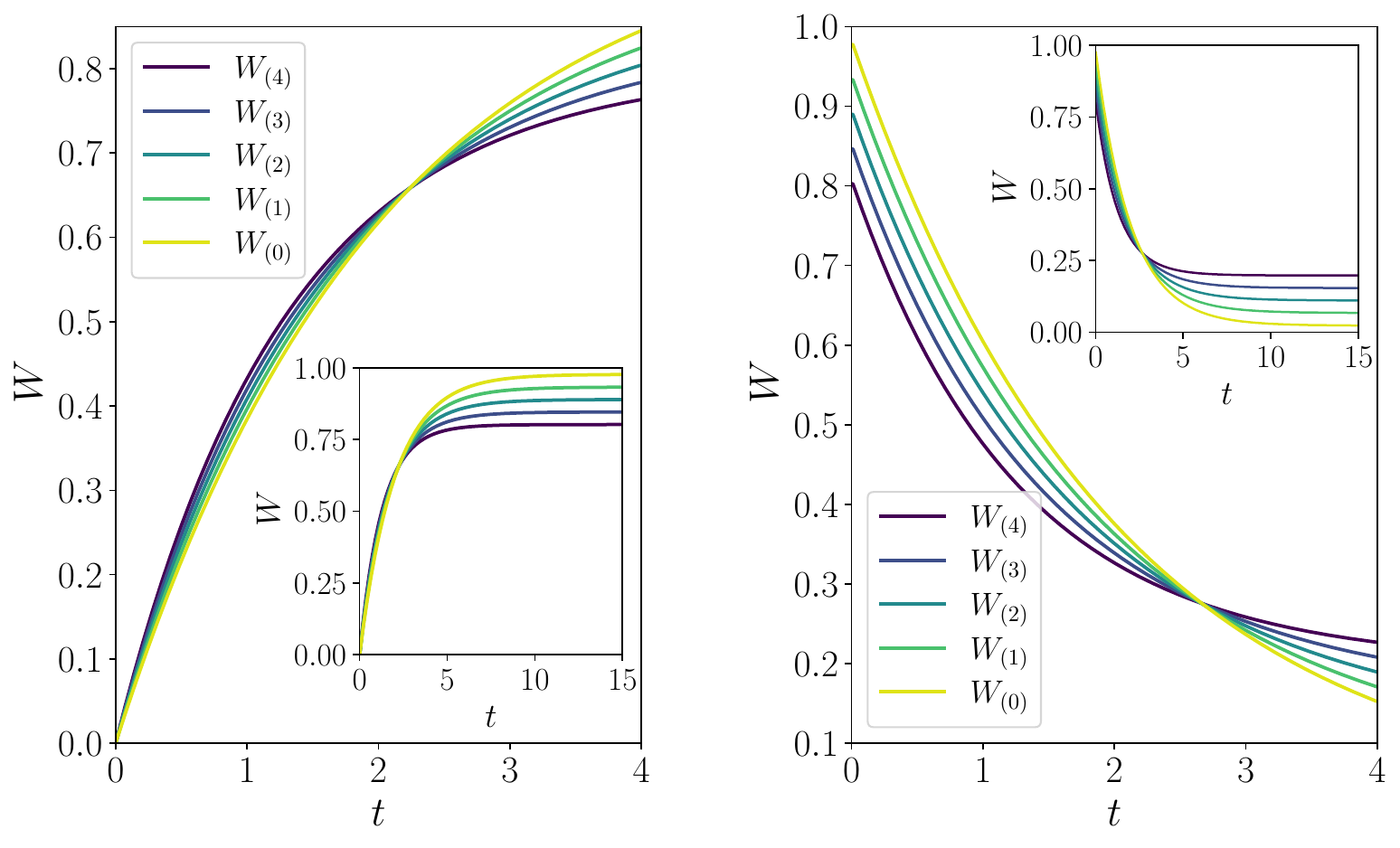}
\caption{(Color online.) Variation of $W$ (ordinate)  against $t$ (abscissa). The initial state is prepared as the ground state of the transverse  Ising chain.
Left panel is for the charging process while the right one is for the discharging case.  Plots are for different number of spins that are exposed to bit-flip noise. $W_{(0)}$ indicates the noiseless case while $W_{(i)},\, i= 1, \ldots,  4$ represents \(i\) number of spins  effected  through dephasing channels. To show the advantage obtained by noise in the transient regime, we plot up to \( t = 4\).  Here $\lambda = 0.5$,  and \(N=4\).  \(\Gamma_{dph}^x/\Gamma_{abs} =0.3\) (left panel)  and \(\Gamma_{dph}^x/\Gamma_{dis} =0.3\) (right panel). Inset: Behavior of $W$ with the variation of  $t$. $t$ is chosen up to the point where all of them goes to a steady state. Both the axes are dimensionless.  } 
\label{fig_work_noise124}
\end{figure}

\begin{figure}
\includegraphics[width=\linewidth]{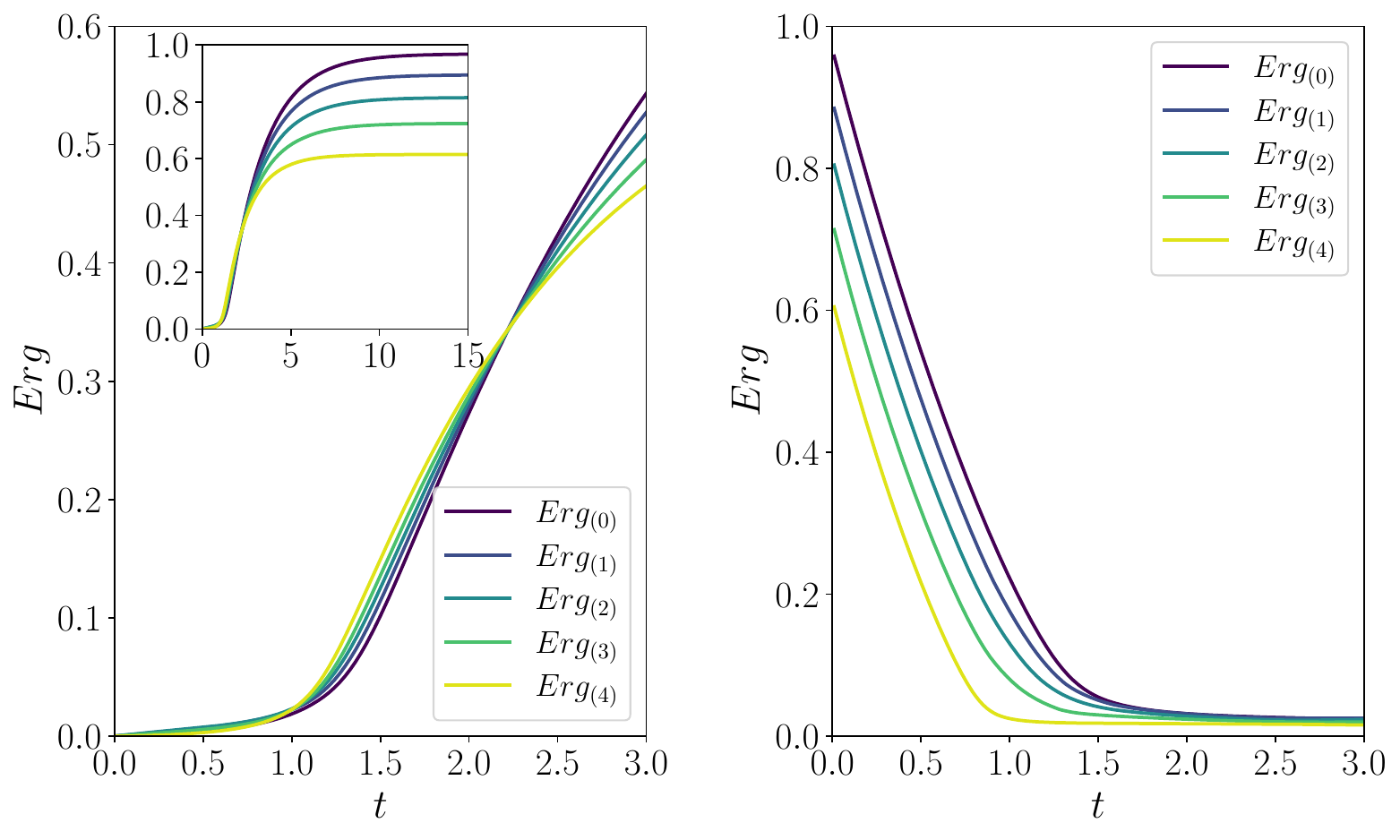} 
\caption{(Color online.) Ergotropy vs. time, $t$. All the other parameters and conditions are same as in Fig. \ref{fig_work_noise124}. Both the axes are dimensionless. }
\label{fig_erg_noise124}
\end{figure}

Let us now consider other  parameter regimes involved in this problem so that we can identify the exact conditions which lead to the gain of noise in the performance of the QB.

\subsubsection{\textbf{Role of noise on charging-discharging.}}  We first demonstrate the effects of local bit-flip noise on QB.  For analyzing this, the ground state of the transverse Ising chain  with \(N=4\) and \(\lambda=0.5\)  is taken as the initial state of the QB. 
To systematically probe the situation, we consider the cases, when  noise acts  
on $i, \ i = 1, \ldots,  4,$ number of spins (marked as $(i), \ i = 1,\ldots, 4$ in Figs. \ref{fig_work_noise124} and \ref{fig_erg_noise124}).
We then compare them with the noiseless scenario which is indicated as $W_{(0)}$ and \(Erg_{(0)}\) in Figs. \ref{fig_work_noise124} and \ref{fig_erg_noise124} respectively. Let us first consider the initial period of time  of the dynamics during charging, i.e., in the transient part of the dynamics.  In this regime, we find a crossover time, \(t_c\), up to which we observe (see Fig. \ref{fig_work_noise124})
\begin{eqnarray}
\label{eq_hierarchyW}
W_{(4)} > W_{(3)} > W_{(2)} > W_{(1)} > W_{(0)},
\end{eqnarray}
where subscripts denotes the number of spins in the QB being influenced with noise.  In Fig. \ref{fig_erg_noise124}, we notice that almost the similar patterns for  ergotropy emerge although in this case, the hierarchy is prominent in the  intermediate time interval instead of initial time-period of the  evolution.  In this case also, there exists \(t< t_c\) upto which  presence of noise actually helps to extract higher amount of work from the battery than in the noiseless scenario  (see Fig. \ref{fig_erg_noise124}).  Both the results for work-output and ergortropy indicate that  for storing energy quickly in the battery described by the transverse Ising model, decoherence outperforms over the noiseless case.  
%This can also be perceived from the behavior of instantaneous power, although with different \(t_c\).
Exactly opposite orderings among optimal work-gained appear when \(t> t_c\) which include the region where the system reaches its steady state.  In particular, in this region,  noiseless scenario gives the maximum storage capacity in the QB. In the saturated cases, energy stored in the battery decreases with the increase of noise  in the system. We will show in the next section that it is possible to overcome  this detrimental picture of a Markovian noisy channel by a non-Markovian one. 
 
 During the extraction of energy,  the absorption channels acted on each spin are replaced by the dissipative ones and local bit-flip noise acts on each spin. 
In this situation, battery discharges rapidly in the noiseless scenario, while in presence of noise, it takes a long time to discharge. Although the behavior of ergotropy remains qualitatively similar in the discharging process, unlike \(W\), no crossover occurs for \(Erg\) between different noisy scenarios. In the entire duration of time, before reaching to the steady state, noise actually helps rapid decay of the extractable work. If we design a device for which quick extraction of energy from the battery is required, dephasing noise turns out to be beneficial as shown in the right panel of Fig. \ref{fig_erg_noise124}.  
%Although the complete extraction of energy from the battery is not possible in case of noisy as well as noiseless scenarios in finite time, 
Note, however, that in absence of noise, total extraction is possible only in the asymptotic case due to a very slow rate of dissipation, i.e.,  the choice of a low value of \(\Gamma_{dis} =0.5\).

\begin{figure}
\includegraphics[width=\linewidth]{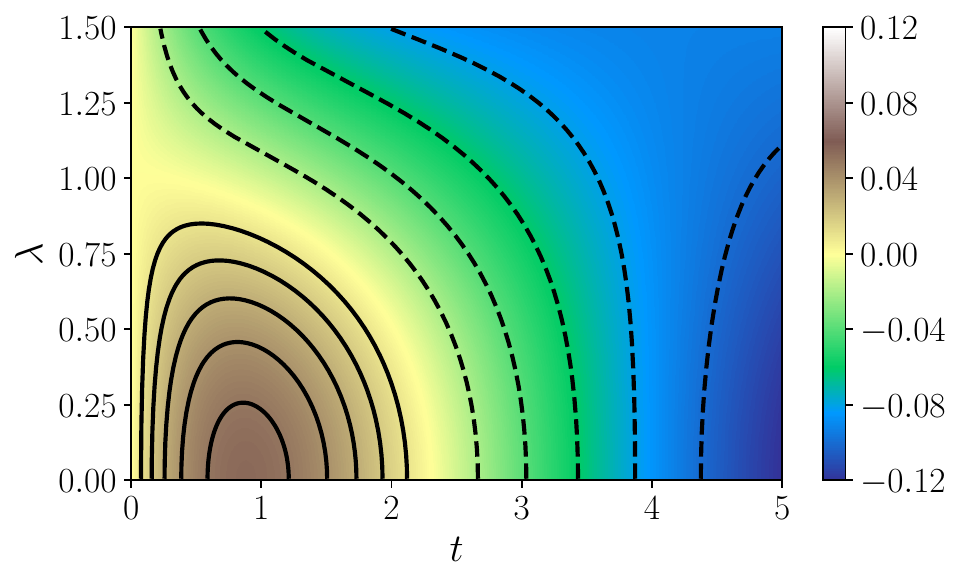}
\caption{(Color online.) Contour plot of  \(\delta_{x>z}^{adv}  =W (\Gamma_{dph}^x) - W (\Gamma_{dph}^z)\) (see text)  with the variation of  $t$ (horizontal axis) and \(\lambda\) (vertical axis). All the other parameters are same as in Fig. \ref{fig_work_noise124}. Both the axes are dimensionless.}
\label{fig_J_noise_variation} 
\end{figure}

\begin{figure}
\includegraphics[width=\linewidth]{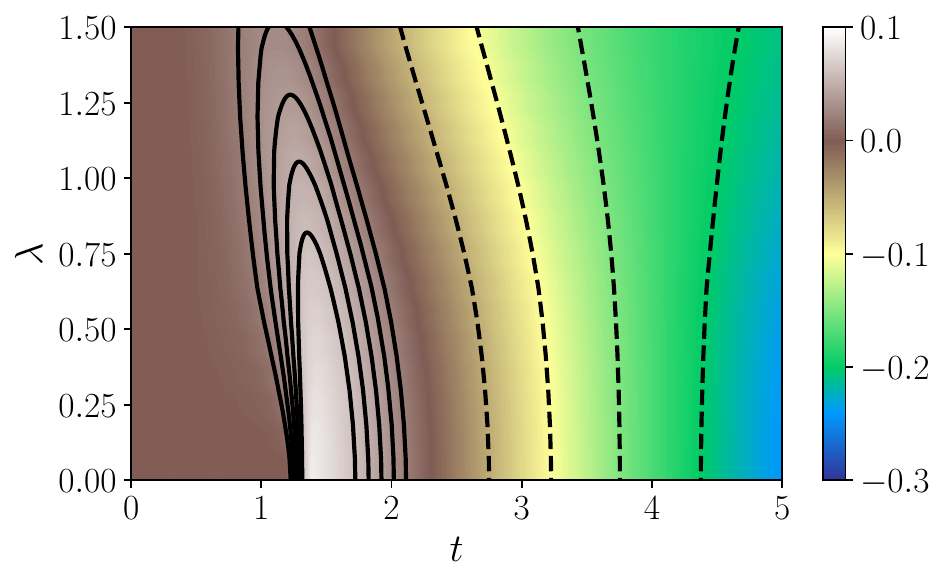}
\caption{(Color online.) Contour plot of  \(\Delta_{x>z}^{adv}  =Erg (\Gamma_{dph}^x) - Erg (\Gamma_{dph}^z)\) (see text) against  $t$ (horizontal axis) and \(\lambda\) (vertical axis). All the other parameters are same as in Fig. \ref{fig_work_noise124}.Both the axes are dimensionless.}
\label{fig_J_noise_variation_ergo} 
\end{figure}

\subsubsection{\textbf{Relation of quantum phases with power of QB.} } All the above discussions were restricted to  local bit-flip noise. As shown in Proposition 3 for the battery with two spins, there exists a range of \(\lambda\) in the paramagnetic regime, where bit-flip noise in all the spins  can produce a higher work-output than that of  the phase-flip ones. Let us now see if we increase the number of spins whether the picture remains same or not. We find that this is indeed the case. 

To make the comparison more effective, we investigate the trends of \(\delta_{x>z}^{adv}  =W (\Gamma_{dph}^x) - W (\Gamma_{dph}^z)\) and \(\Delta_{x>z}^{adv}  =Erg (\Gamma_{dph}^x) - Erg (\Gamma_{dph}^z)\) with time and the coupling constant \(\lambda\) of the QB. Note that since Markovian noise can give advantage only in the transient regime, we are interested only in that period. Fig. \ref{fig_J_noise_variation} also depicts that  the dependence of time and \(\lambda\) in \(\delta_{x>z}^{adv} \) is complementary in nature. On the other hand, intermediate time of the evolution is clearly beneficial as far as ergotropy is concerned as depicted in Fig. \ref{fig_J_noise_variation_ergo}.  Let us summarize all the observations for \(\delta_{x>z}^{adv}\) and \(\Delta_{x>z}^{adv}\)  in this respect.

\emph{Observation 1.}  For small values of \(\lambda\), i.e., when the transverse Ising model is deep into the paramagnetic phase, the bit-flip noise is always better than the phase-flip one with respect to storing of energy.  \\
\emph{Observation 2.} With the increase of \(\lambda\),  the time up  to which \(\delta_{x>z}^{adv}\) remains positive decreases. For example, with \(\lambda =0.5\), it is positive up to \(t \approx 2.01\) while \(\lambda=0.9\), the time reduces to    \( \approx 1.0\). \\
\emph{Observation 3.}  If   QB is prepared in the antiferromagnetic phase at \(t =0\), the phase-flip noise can give more work-output during charging than the bit-flip one in the transient regime.\\
However, by considering the figure of merit as ergotropy, we find out the following:\\

\emph{Observation 1.}  Fig. \ref{fig_J_noise_variation_ergo} clearly  demonstrates that  in the intermediate time of the dynamics, bit flip noise is helpful to obtain high ergotropy   in comparison with phase-flip one. \\
\emph{Observation 2.} %At a particular value $t$, 
The behavior of $\Delta_{x>z}^{adv}$ depends almost negligibly on $\lambda$ as seen from the almost parallel lines with \(\lambda\)  in Fig. \ref{fig_J_noise_variation_ergo}, especially for high values of \(t\), i.e., the trends of ergotropy do not crucially depend  on the phase of the transverse Ising model. This behavior is in sharp contrast with stored work difference $\delta_{x>z}^{adv}$ in Fig. \ref{fig_J_noise_variation}.

\subsubsection{\textbf{Dependence on anisotropy.} } Let us justify here the reason to choose \(\gamma =1\) i.e., transverse Ising model for presenting all the results. When there is no noise in the system, we can find that the storage of work can increase with the decrease of the anisotropy in the  $XY$ model and similarly more extraction of energy is possible for low values of \(\gamma\). Such a difference  wipes out when the noise acts on the system, i.e., there is almost no difference found in the trends of \(W\) for different values of \(\gamma\) in presence of noise, thereby confirming that the results reported here also hold for other values of \(\gamma\).

%\begin{figure}
%\includegraphics[height=5.0cm,width=9.0cm]{scale_inv_n2_n4.pdf}
%\caption{(Color online.) $\mathcal{P}_{ins}$ (ordinate) vs. $t$ (abcissa). Instantaneous power is computed in different times for different system size $N$. Here $J = 0.5$. Here we consider markovian dynamics of the system. Figure in the left is for absorption cycle and right one is for dissipation cycle. We follow the same left-right combination for rest of the paper, unless stated otherwise.}
%\label{fig_scale} 
%\end{figure}

\subsubsection{\textbf{Scale invariance.} }  We can check how the  results scales with the increase of the system-size of the battery. We notice that after normalizing the Hamiltonian, as in Eq. \eqref{eq:normalizing}, the work-gained (or -extracted)  or the ergotropy  has no qualitative as well as quantitative change with the increase in the number of spins in the battery.   
%\textcolor{blue}{IS it true? ergotropy is scale invariance?}
%Considering bit-flip noise on each qubits, 
%in the left panel of Fig . \ref{fig_scale},  
%we apply a single-mode bosonic absorption channel in each of the qubit for time interval $t = 0$ to $t = 15$  while in the right panel,   dissipative channels act  on each qubit for rest of the time period and in both the scenarios, 
For example,  \(W\) for \(N=2\) and  \(N=8\)   are essentially same up to the numerical accuracy of \(10^{-4}\). 
%We also check that such a feature persists even when \(N>4\). 
Moreover,  a good agreement of  the above results  for  \(N=4\)  with the Propositions  in Sec. \ref{subsec_QBmarko2} also confirms the scale-invariance, thereby indicating the validity of all the results for arbitrary number of spins of the battery.

\begin{figure}
\includegraphics[width=0.7\linewidth]{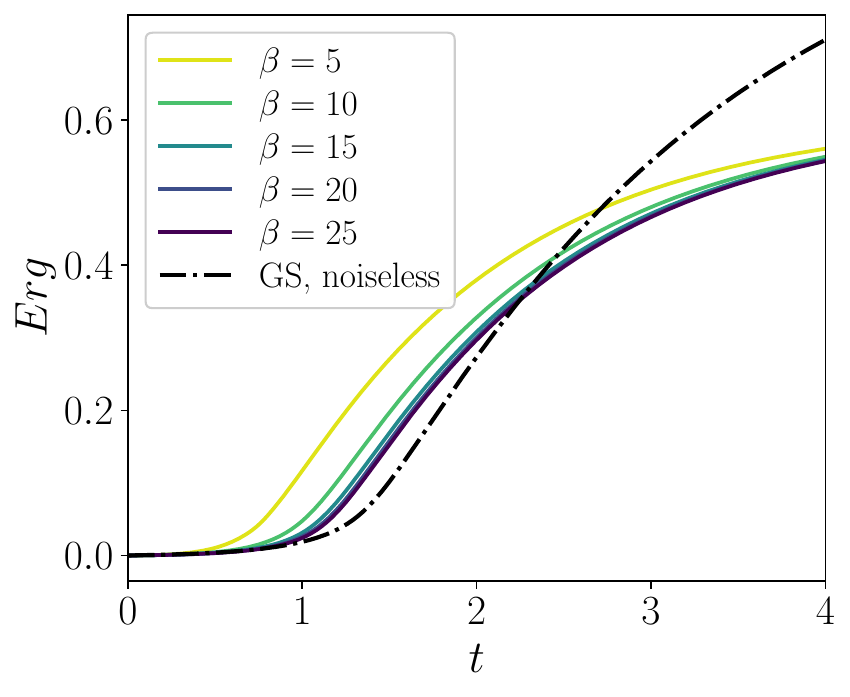}
\caption{(Color online.)  $Erg$ (ordinate) against $t$ (abscissa) for canonical equilibrium state of the transverse Ising model as the initial state of the battery. Different  (solid) lines correspond to different values of $\beta$ in presence of Markovian noise, and from top to bottom, \(\beta\) is decreasing. Dot-dashed line represents the noiseless case with the ground state being the initial state of the battery.  We only consider when the absorption channel is connected to QB, i.e., in the charging process.  Other parameters are similar to Fig. \ref{fig_work_noise124}. Both the axes are dimensionless.}
\label{fig_thermal} 
\end{figure}

\subsubsection{ \textbf{Finite temperature tolerance of QB under dephasing noise. }} Up to now, we  prepare  the ground state of the transverse Ising  Hamiltonian as the initial state of the battery.
 From an experimental point of view, it is not  possible to reach the absolute zero temperature. Therefore, it is interesting to check whether the noise-induced performance of QB constructed via transverse Ising model persists even in a finite temperature or not. 
To investigate it,  let us take the initial state as the thermal state of the transverse Ising spin chain  which can be represented as  $\rho_{\beta} = \exp(-\beta H_B)/Z$, with $\beta = \frac{1}{k_{B}T}$, $T$ being the  temperature, and $k_{B}$ the Boltzmann constant. Here $Z = \mbox{Tr} (\exp(-\beta H_B))$ is the partition function of the  system. Keeping all other parameters unchanged, and by applying dephasing channels in all the spins in the \(x\)-direction,  we study  the variation of ergotropy with respect to time for different values of inverse temperature $\beta$. In Fig. \ref{fig_thermal}, we show that again at the intermediate time period in the transient regime,  noisy states are beneficial in terms of  ergotropy, even at finite temperature.  Specifically, in this noisy situation,  the value of ergotropy  increases   with the decrease  of $\beta$, i.e., with the increase of temperature which establishes the robustness of the quantum battery based on quantum spin chain. Moreover, in that period of time, the ground state as the initial state without noise performs worst in terms of ergotropy production than that of any other thermal states with noise. 
%In others, Fig. \ref{fig_thermal} efficiently depicts that for low values of $\beta$, i.e.,  for high temperature along with the noise,  there exists a particular time interval where the value of ergotropy is higher than that of the  ground state without noise. 

%\sout{We notice that  in the transient regime, beneficial behavior of noisy scenarios over a noiseless ones remains valid even for moderately high temperature.  In Fig. \ref{fig_thermal}, different lines correspond to  $\mathcal{P}_{ins}$ with different values of $\beta$. 
%So, as we investigate earlier that, in case of charging of a quantum battery, the advantageous behaviour persists even in thermal state as initial state also, which is experimentally more easier to implement.
%For example, we observe that  there always exist \(t_c\) below which such an advantage under local Markovian bit-flip noise  can be found for \(\beta \gtrsim 9.4\)  with \(\lambda=0.5\), and \(\Gamma_{dph}^x/\Gamma_{abs} =0.3\) up to numerical accuracy (see Fig. \ref{fig_thermal}).} 

\subsubsection{\textbf{Entanglement in dynamics.}} We compute the nearest neighbor entanglement, quantified by logarithmic negativity \cite{vidal'02},  between spin \(1\) and \(2\) after tracing out the rest of the spins. We find that although the initial state of the tranverse Ising model acted as QB is entangled \cite{fazioreview, aditireview}, the entanglement of the evolved state  decreases with the increases of time and finally vanishes after a short period of time, both in the noiseless and in the noisy scenarios. Moreover, in the transient regime, entanglement  of the noisy state is less than that of the noiseless case. Hence in this model,  generation of entanglement in the charging process is not important to achieve better performances of QB  (cf. \cite{munro'20}).

%\begin{figure}
%\includegraphics[height=5.0cm,width=7.0cm]{r_variation_sigmaz.pdf}
%\caption{•}
%\label{•} 
%\end{figure}

%\end{enumerate}
%\noindent

\section{Non-Markovian noise leads to better efficiency in quantum battery}
\label{sec_QB_nonMarko4}

\begin{figure}
\includegraphics[width=\linewidth]{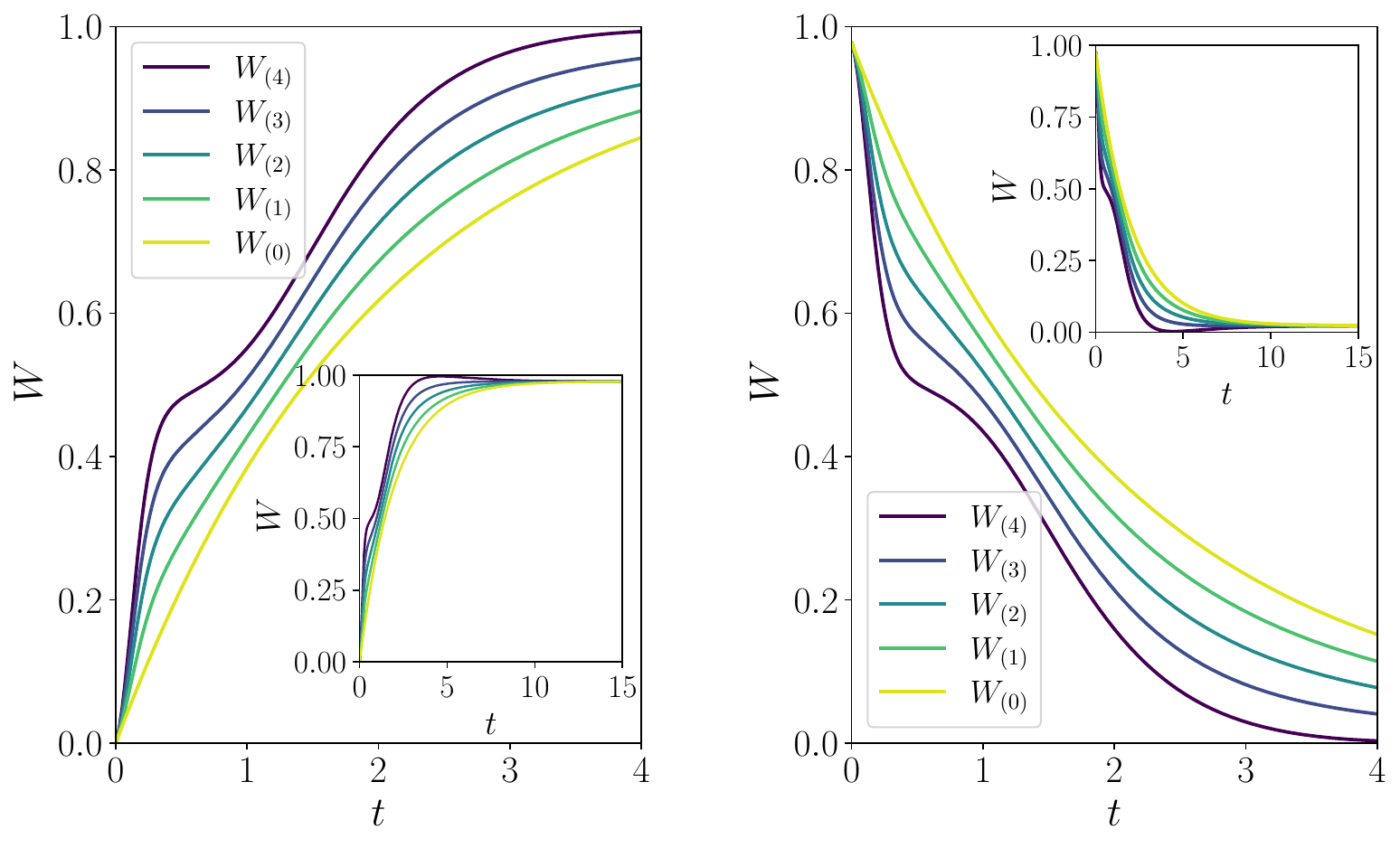}
\caption{(Color online.) Pattern of work (y-axis)  vs. time (x-axis).  Left and right panels are respectively for charging and discharging methods.  Here, Ohimicity parameter is chosen as $s = 4$, $\lambda = 0.5$, and \(\Gamma_{abs} = \Gamma_{dis} = 0.5\). Comparing the above work-output with the one in Fig. \ref{fig_work_noise124}, we clearly see that non-Markovianity induces higher storage and extraction capacities of QB  than that of the Markovian noise.  All other specifications are same as in Fig. \ref{fig_work_noise124}. Both the axes are dimensionless.  } 
\label{fig_nonmarkoW}
\end{figure}

In the preceding section, we restrict to the time-independent strength of the Lindblad operators that leads to the Markovian dynamics of the reduced battery-system. If this condition is relaxed,  and we now want to see the memory effect in the charging-discharging cycle of the battery, 
% of the system 
we consider the evolution of the battery to be non-Markovian which was shown to increase several  resources like entanglement, coherence, of quantum states, thereby ensuring the possibility to increase the efficiency in quantum information processing tasks \cite{laine'10, rivas'10, lu'10, haikka'13, Chin'12, vasile'11, thorwart'09, huelga'12, schmidt'11, sabrinarev, titassamya, rivu'20}. These results motivate us to study the dynamics of QB in this context.

 \subsubsection{\textbf{The QB model in non-Markovian domain}} The system still remains the quantum XY model with a transverse magnetic field. 
% If we allow the back flow of information from the environment to the system then the presence of memory effect is ensured by non-markovian dynamics. In this section we want to point out the effect of non-markovianity in the charging power of our quantum battery which we consider as an one-dimensional ising spin chain in open boundary condition. \\
Again each spin of the battery  is  connected to bosonic reservoirs for storing and for extracting energy. Each spin  of the battery is further attached to the dephasing channel which can be either in the \(x\)- or in the \(z\)-direction. Unlike Markovian channel, we assume that  the strength of the dephasing channel, \(\gamma_{(3)}\), is time-dependent, i.e., \(\Gamma_{dph}^x(t) \) (\(\Gamma_{dph}^z(t) \))  can be characterized by the Ohmic spectral density \(G(\omega)\) \cite{haikka'13}, given by 
\begin{equation}
G(\omega) = \frac{\omega ^{s}}{\omega^{s-1}_c} \exp(-\frac{\omega}{\omega_{c}}),
\end{equation}  
with \(\omega\)  and \(\omega_c\) being  respectively the frequency and the cut-off frequency  of the reservoir and \(s\) is the Ohmicity parameter. 
%.  to dephasing e a thermal reservoir. The GSKL master equation corresponding to the Hamiltonian then take the form,
%
%\begin{equation}
%\frac{d\rho}{dt} = \Gamma_{deph}(t)(\sigma_z \rho(t) \sigma_z - \rho(t)),
%\end{equation}
%Here, unlike the previous case of markovianity, $\Gamma_{deph}(t)$ represents the time-dependent dephasing rate which is chrecterized by the the spectral density $G(\omega)$ of the thermal reservior that is connected to each of the qubit. $\omega$ is the frequency of the reservior. Now considering the initial state of the environment as thermal one, the dephasing rate takes the form,
%Each qubit can be  affected by thermal reservoir
%\begin{equation}
%\Gamma_{deph}(t) = \int^{\infty}_{0} d\omega G(\omega) coth[\omega / 2k_{B}T]sin(\omega t)/ \omega,
%\end{equation}
%where $T$ is the absolute temperature of the reservior and $k_{B}$ is the Boltzman constant.
%For our paper we adopt ohomic spectral density function, which is of the form,
%
%
%
%Here, $\omega_{c}$ is the cut-off frequency and $s$ is the ohimicity parameter. Equipped with this, in zero temperature limit Eqn. (26) takes the form 
The noise strength reads as 
\begin{equation}
\Gamma_{deph}(t,s) = (1+ (\omega_{c}t)^{2})^ {- \frac{s}{2}} \Gamma(s) \sin[s \tan^{-1}(\omega_{c} t)],
\end{equation}
where $\Gamma(s)$ is the Euler gamma function.  It can be  checked that  when $s > 2$, the above mentioned dephasing rate can become temporarily negative and memory effect assists the back-flow of information from the environment  to the QB, thereby representing a non-Markovian domain, while \(s\leq 2\) is the Markovian regime \cite{haikka'13}. Importantly,   such an environment was shown to  be realized by ultracold atoms \cite{worknew}.

%. So, the system responds on the non-markovian dynamics only when $s > 2$ i.e., in the super-Ohomic region of the reservior spectrum.

\subsection{Non-Markovian dynamics always enhances the performance of QB}

\begin{figure}
\includegraphics[width=\linewidth]{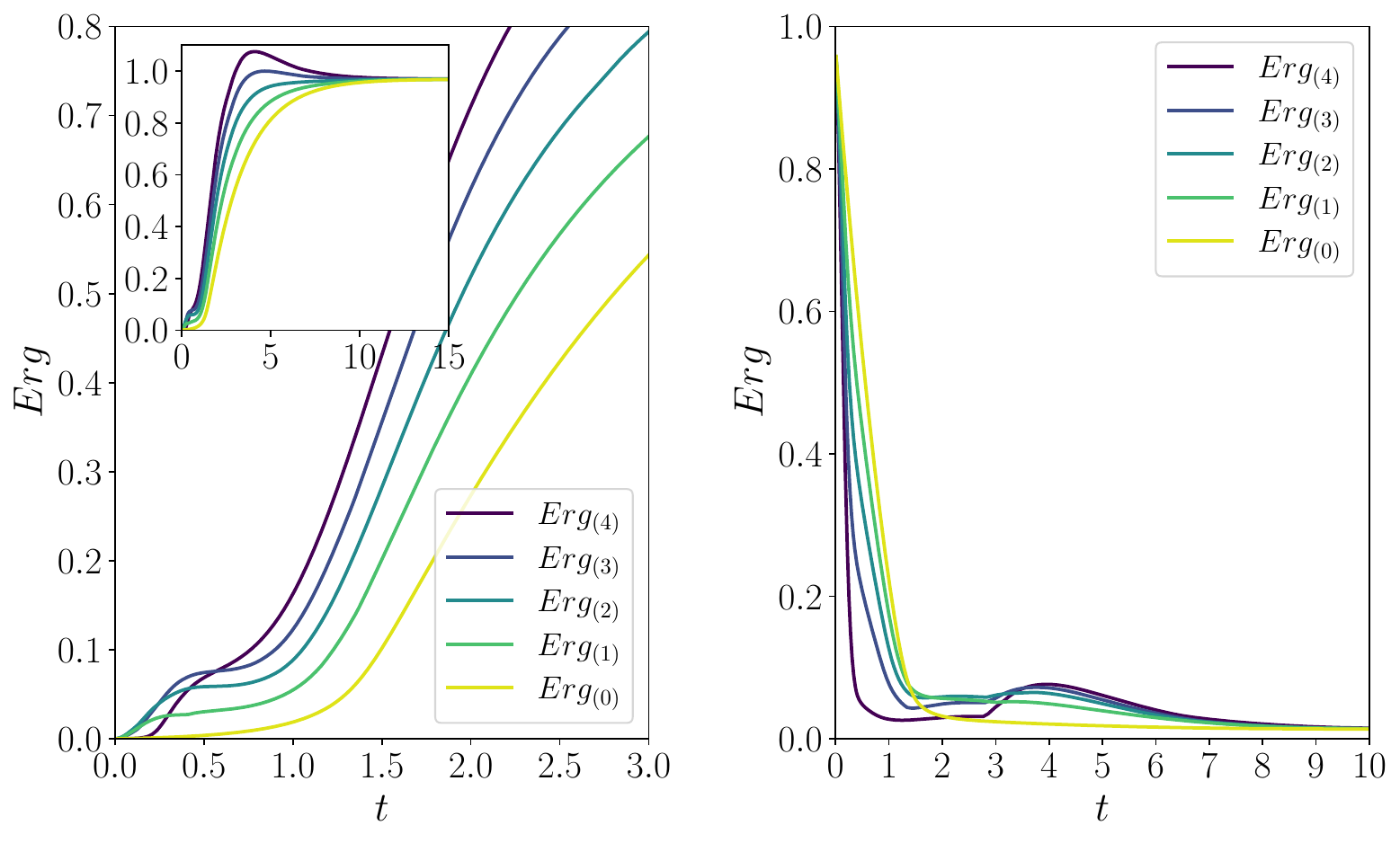}
\caption{(Color online.) Trends of $Erg$ (ordinate)  with time (abscissa). One should compare the behavior of $Erg$  in the non-Markovian domain with the Markovian one in Fig. \ref{fig_erg_noise124}. All other parameters are same as in Fig. \ref{fig_nonmarkoW}. Both the axes are dimensionless. }
\label{fig_noise124_nonmarko} 
\end{figure}

\subsubsection{\textbf{Charging-discharging process.} } Like Markovian case, the system is initially prepared in the ground of the transverse Ising model, which is attached through a bosonic reservoir with \(\Gamma_{abs} = 0.5\). In case of decoherence,  we again compare \emph{five} situations -- when there is no noise in the system, and when  a single spin, or two, or  three, or all the spins are affected by bit-flip noise. With a sharp contrast to the Markovian noise, we report here that  in the charging  process, the amount of work stored increases with the increase of noise in the system, both in the  transient as well as the steady-state regime and interestingly, the same feature persists in case of ergotropy as depicted in Figs. \ref{fig_nonmarkoW} and  \ref{fig_noise124_nonmarko}. Specifically, we  find that for extractable work, there is no existence of crossover time, as obtained in the Markovian scenario  and in the entire evolution-time, \(t\), we have
\begin{eqnarray}
\label{eq_nonMarkohierarchyW}
W_{(4)} > W_{(3)} > W_{(2)} > W_{(1)} > W_{(0)}, \,\, \forall t
\end{eqnarray}
where the notations are the same as in Eq. (\ref{eq_hierarchyW}) as depicted in the left panels of Fig. \ref{fig_nonmarkoW}  while the similar hierarchy is also observed for ergotropy after an initial time interval  as shown in Fig. \ref{fig_noise124_nonmarko}. It clearly shows that the non-Markovian noise in all the four spins can lead to higher storing capacity  of energy (ergotropy) in QB of transverse Ising model compared to the low noise and noiseless cases, thereby showing the usefulness to build noise-induced QB.  

The discharging process also shows a remarkable improvement in presence of non-Markovian noise --  extracted amount of energy and ergotropy increases with the increase of noise as seen in the right panels of Figs. \ref{fig_nonmarkoW} and \ref{fig_noise124_nonmarko} with \(\Gamma_{dis} = 0.5\). Moreover, strong effect of back-flow of information can be perceived from the patterns of work-gained and -extracted (Fig. \ref{fig_nonmarkoW}). Unlike Markovian regime, complete extraction of energy from the battery is possible when all the spins are under local bit-flip noise.

\begin{figure}
\includegraphics[width=\linewidth]{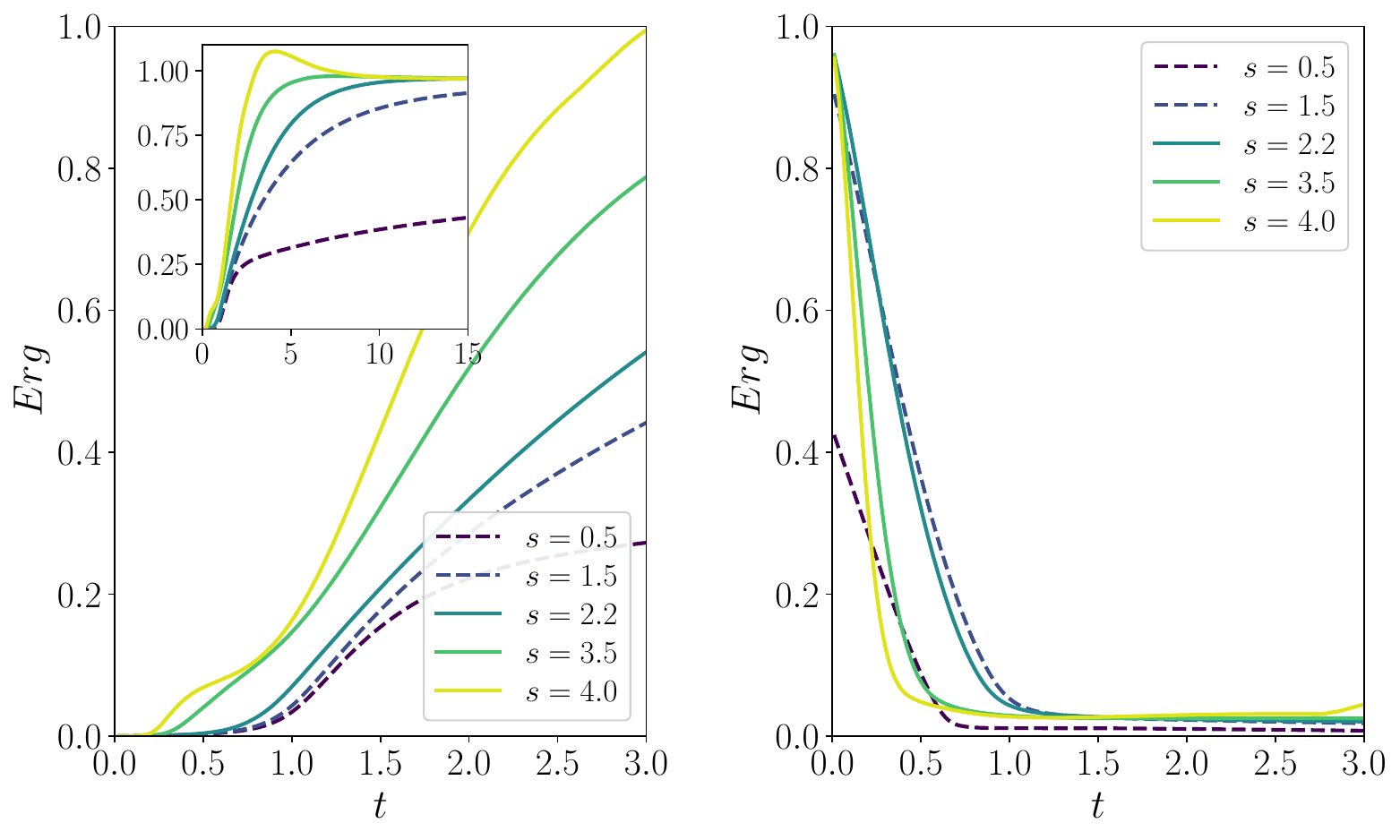}
\caption{(Color online.) Behavior of $Erg$ (vertical axis)  against \(t\) (horizontal axis) for charging (left panel) and discharging (right panel) processes. Different lines represent maximal work-gained (-extracted) i.e., ergotropy for different Ohmicity parameter, \(s\). Dashed lines are for \(s=0.5\) and \(1.5\) representing Markovian dynamics while the rest  (solid) lines indicate the non-Markovian ones.   From figure, we clearly see the advantage that one  can gain with the introduction of non-Markovianity.   Other choices of parameters are same as in Fig. \ref{fig_nonmarkoW}. Both the axes are dimensionless.}  
\label{fig_markov_nonmarkov} 
\end{figure}

\subsubsection{\textbf{Markovian vs. non-Markovian dynamics.} } We have already seen that there is a clear difference between Markovian and non-Markovian noise. To perform the comparison more concrete, we now investigate  the patterns of work-output as well as ergotropy  by changing the Ohmicity parameter, \(s\), from \(0.5\) to \(4\), i.e., by sweeping the system from Markovian to non-Markovian domain, when all the spins which are initially prepared as the ground state of the transverse Ising chain are affected by local noise.   Firstly,  in the charging scenario,   \(Erg\) as well as \(W\)  increases faster with the increase of \(s\) and they  reach their  maxima   with \(s =4\). Quantitative analysis reveals that the difference between \(Erg\) with \(s =0.5\) and \(s=4.0\)  is of the order of \(\approx 0.537\) when both of them saturates (see Fig. \ref{fig_markov_nonmarkov}).
 The second observation  is the nonmonotonic nature of \(Erg\) (\(W\)) in the non-Markovian case which is absent in the Markovian domain (Fig. \ref{fig_markov_nonmarkov}). In case of discharging, noise induces the fast and high amount of extraction of energy as well as ergotropy  from the battery.  
 %and again the work-extracted monotonically increases  with the increase of \(s\).  
 We make the other observations below by fixing the Ohmicity parameter to be \(4\). 
 % \textcolor{blue}{s=0.5 behave strangely, please check, why it starts from 0.4, work extracted to monotonic noi, then? }

\begin{figure}
\includegraphics[width=\linewidth]{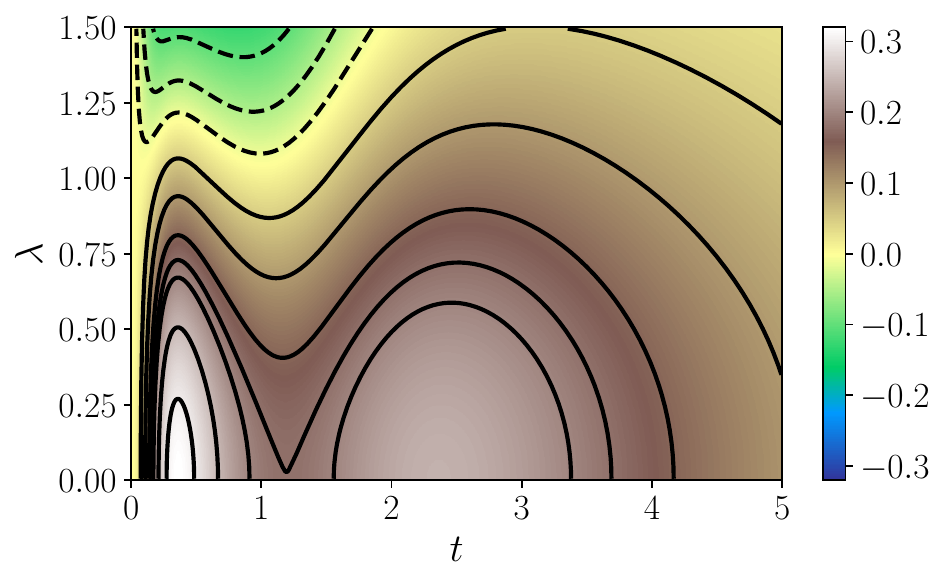}
\caption{(Color online.) Contour plot of $W(\Gamma_{dph}^x) - W (\Gamma_{dph}^z)$ during the charging process with the variation of time (x-axis) and \(\lambda\) (y-axis). The similar dependence can be found for the Markovian case in Fig. \ref{fig_J_noise_variation}. All other specifications are same as in Fig. \ref{fig_nonmarkoW}. Both the axes are dimensionless.}
\label{fig_J_noise_variation_nonmarko} 
\end{figure}

\begin{figure}
\includegraphics[width=\linewidth]{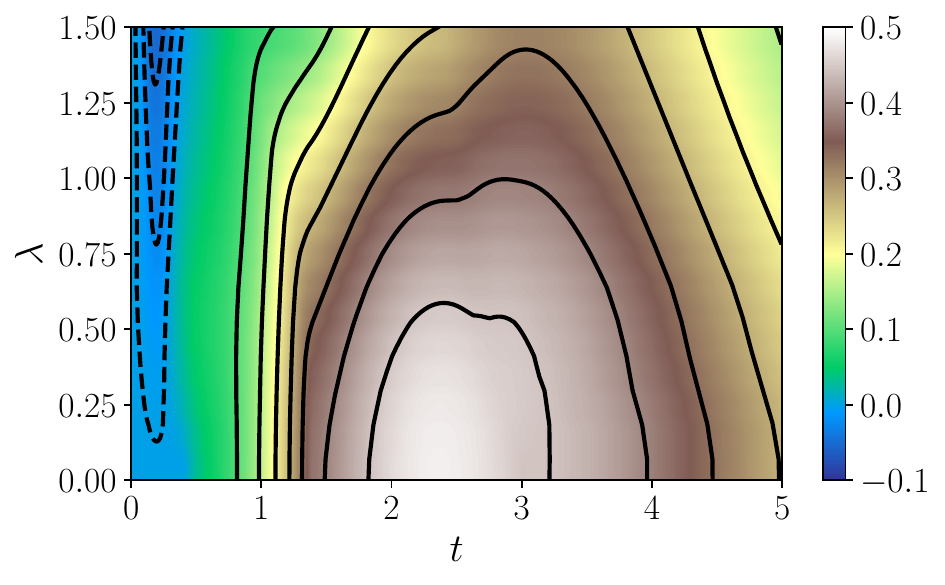}
\caption{(Color online.) Contour plot of $Erg(\Gamma_{dph}^x) - Erg (\Gamma_{dph}^z)$ with time (x-axis) and \(\lambda\) (y-axis). All other specifications are same as in Fig. \ref{fig_nonmarkoW}. Both the axes are dimensionless.}
\label{fig_J_noise_variation_nonmarko_ergo} 
\end{figure}

\subsubsection{\textbf{Dephasing noise vs. interaction strength.} } To study the role of the direction of noise on the work-stored and -extracted, we plot \(\delta_{x>z}^{adv}  =W (\Gamma_{dph}^x) - W (\Gamma_{dph}^z)\)
in Fig. \ref{fig_J_noise_variation_nonmarko} with the variation of  time and \(\lambda\) (cf.  Fig. \ref{fig_J_noise_variation}) during charging of the battery. Comparing Fig. \ref{fig_J_noise_variation} with Fig. \ref{fig_J_noise_variation_nonmarko}, we find a  difference between Markovian and non-Markovian cases due to the nonmonotonic nature of   work-output in the later. Surely, if the initial state belongs to  the paramagnetic regime, unlike the Markovian case,  the advantage  of local bit-flip noise over the phase-flip one, quantified by  \(\delta_{x>z}^{adv}\), can be of the order of \(\approx 0.05 - 0.3\). On the other hand, in Fig. \ref{fig_J_noise_variation_nonmarko_ergo}, we observe a striking difference in ergotropy compared to that obtained in Fig. \ref{fig_J_noise_variation_ergo} for Markovian noise. In the non-Markovian scenario, we report $\Delta_{x>z}^{adv} \geq 0 $ after the transient regime, i.e.,  the bit-flip noise is always better than the phase-flip one except for small values of \(t\). Specifically,  a significant amount of  advantage is observed for intermediate time, $t \approx (2,3)$ and for small values of $\lambda$.

%we observe that if you put the interaction strength of the qubits in the paramagnetic case as in the case of markovian one we will get advantageous behaviour from our battery. \emph{Direction of noise is responsible for high power :} Same as markovian case. If we put the direction of noise along x direction we will get positive outcome from our system. 

 Like the Markovian case,  we also observe  that system-size does not play a role (after normalizing the Hamiltonian) in the storing-extracting process of QB  in  non-Markovian regime, thereby again confirming  the scale-invariance feature. Moreover, we notice that bipartite entanglement between spin \(1\) and \(2\) in presence of non-Markovian noise collapses,  then again revives during charging as well as discharging and the amount of entanglement increases with the increase of Ohmicity parameter. However, the pattern of entanglement cannot explain the behavior of work-output as well as ergotropy. In particular, when the bipartite state is separable, we can still obtain stored energy or extracted energy  (maximal extractable work) to be greater in case of non-Markovian noise than the noiseless and Markovian ones. 
 %In Fig. \ref{scale_nonmarko} it is clear that there is no qualitative as well as quantitative changes even if we allow information back flow from environment to our battery.\\

%\begin{figure}
%\includegraphics[height=5.0cm,width=8.5cm]{scale_inv_n2_n4_nonmarkovian.pdf}
%\caption{(Color online.)Dependance of instantaneous power on system size $N$, for $J = 0.5$ . We plot $\mathcal{P}_{ins}$ with respect to $t$ for different system size in case of nonmarkovian dynamics.}
%\label{scale_nonmarko} 
%\end{figure}

\section{Discussion}
\label{sec_conclu}

Preparing an isolated system is essentially impossible in laboratories and hence decoherence is one of the main obstructions which reduce the performance of the devices. The main reason behind such observations is that all the quantum properties like entanglement, coherence responsible for quantum advantage are fragile in the presence of noise. Therefore,  designing quantum machinery which is robust against noise has immense importance in quantum technologies. We address this question in the context of a storage device, especially for the quantum battery.  

We considered  the ground or the canonical equilibrium state of an interacting spin-1/2 chain, specifically a quantum $XY$ model with a transverse magnetic field as the initial state of the quantum battery (QB) which is exposed to an environment. For storing  energy, each spin-1/2 particles of the battery are connected to bosonic reservoir which acts as  a charger. During the extraction of energy from the battery, another set of bosonic reservoirs are attached to the system, which helps to dissipate the energy. Moreover, some or all the spins are exposed to a local noisy environment which is modelled as   bit- and phase-flip Markovian as well as non-Markovian noise. Our aim was to find the consequence of noise on the performance of the battery. 

In general, we know that decoherence decreases the physical property of the system, thereby reducing the efficiency of the devices. We usually see some exceptions when the noise is of non-Markovian kind due to the memory effects. We reported here that even in presence of Markovian, phase- and bit-flip noise, storing of energy in the battery as well as maximum  work-gained quantified via ergotropy can be enhanced in the transient regime while during the extraction of work from the battery, the process can be made faster in presence of noise. The exact condition for obtaining such improvement in case of work-output is analytically found when QB consists of two spins and was argued to be true for ergotropy. We then showed that such increment persists even with more number of spins in the battery.  We found that  any initial canonical equilibrium state under decoherence can outperform over the noiseless scenario in the transient regime, although the ground state of the transverse Ising model as QB gives the maximum advantage in the stead state.  From numerical simulations,   we identified the proper  range in the parameter-space where such noise-induced work and ergotropy from the QB can be obtained.  We observed that the entire charging-discharging process can be upgraded in presence of non-Markovian Ohmic noise for the entire time period. Therefore, both in the transient and the steady-state regime, non-Markovian noise can perform better than the Markovian as well as noiseless cases. Since both the noise models considered here can be realized in cold-atomic systems, we believe that such  improvements in quantum battery can be achieved in the laboratory. 

\acknowledgements
We  acknowledge the use of \href{https://github.com/titaschanda/QIClib}{QIClib} -- a modern C++ library for general purpose quantum information processing and quantum computing \cite{titas} 
 TC acknowledges support of the National Science Centre (Poland) via QuantERA programme No. 2017/25/Z/ST2/03029.

\appendix

\section{Master equation: Dephasing Noise acted on two spins of QB}
\label{sec_appendix}
Let us first consider the case when phase-flip noise acts on both the spins independently. 
Starting with the initial state as the ground state in Eq. (\ref{eq_gr}), we have to solve Eq. (\ref{eq_master2q}) to get the time-evolved two-qubit state, \(\rho_t\) at any time instant \(t\). To obtain the solution, we work with the  following coupled liner differential equations, given by
\begin{eqnarray}
\dot{\rho_{11}} &=& \Gamma_{abs} (\rho_{22} + \rho_{33}) - \frac{J}{2}\rho_{14}, \nonumber\\
\dot{\rho_{22}} &= &\Gamma_{abs} (-\rho_{22} + \rho_{44}) - \frac{J}{2}\rho_{23}, \nonumber\\
\dot{\rho_{33}} &= &\Gamma_{abs} (-\rho_{33} + \rho_{44}) + \frac{J}{2}\rho_{23},\nonumber\\
\dot{\rho_{44}} &= & -2 \Gamma_{abs} \rho_{44} + \frac{J}{2}\rho_{14}.
\end{eqnarray}
\begin{eqnarray} 
\dot{\rho_{12}} &=& -2 \Gamma_{dph}^z \rho_{12} + \Gamma_{abs} (-\frac{\rho_{12}}{2}+\rho_{34}) + \mbox{Im}, \nonumber\\
\dot{\rho_{13}} &=& -2 \Gamma_{dph}^z \rho_{13} + \Gamma_{abs} (-\frac{\rho_{13}}{2}+\rho_{24}) +\mbox{Im},\nonumber \\
\dot{\rho_{14}} &=& (-\Gamma_{abs} - 4 \Gamma_{dph}^z) \rho_{14} + \mbox{Im},\nonumber\\ 
\dot{\rho_{23}} &=& (-\Gamma_{abs} - 4 \Gamma_{dph}^z) \rho_{23} +\mbox{Im}, \nonumber\\ 
\dot{\rho_{24}}& =& (-\frac{3\Gamma_{abs}}{2} - 2 \Gamma_{deph}) \rho_{24} +\mbox{Im}, \nonumber \\ 
\dot{\rho_{34}}& =&  (-\frac{3\Gamma_{abs}}{2} - 2 \Gamma_{dph}^z) \rho_{34} +\mbox{Im}.\nonumber\\
\end{eqnarray} 
Here \(\rho_{ij}, \, \, i,j =1\ldots 4\) are the matrix elements of the \(4\times 4\) matrix of \(\rho_t\) and \(\mbox{Im}\) is the imaginary part of the matrix elements.  Interestingly, after solving differential equations, all imaginary parts vanish. Multiplying \(\rho_t\) with \(H_B\) in Eq. (\ref{eq_HBsimple}), only a few nonvanishing terms survive and hence we can obtain \(W(\Gamma_{dph}^z)\).

%\section{Bit-flip noise on two qubits of QB}

When dephasing noise in the the $x$-direction acts individually on each spins,   the master equation can be written as
\small
\begin{eqnarray} 
\label{eq_masterbitf}
\nonumber
\frac{d\rho}{dt} &=& -i[H_{B},\rho(t)] \nonumber \\
&+& \Gamma_{abs}[(\sigma^{+}\otimes I) \rho(t) (\sigma^{-}\otimes I) - \frac{1}{2} \lbrace (\sigma^{-}\otimes I) (\sigma^{+}\otimes I) , \rho(t) \rbrace  \nonumber \\
&+& (I \otimes \sigma^{+}) \rho(t) (I \otimes \sigma^{-}) - \frac{1}{2} \lbrace ( I\otimes \sigma^{-}) ( I\otimes\sigma^{+}) , \rho(t)\rbrace] \nonumber \\
&+& \Gamma_{deph}^x [(\sigma_{x} \otimes I) \rho(t)(\sigma_{x} \otimes I) + (I \otimes \sigma_{x}) \rho(t) (I \otimes \sigma_{x})  - 2\rho(t) ]. \nonumber \\
\end{eqnarray}
\normalsize
Taking ground state as the initial state, the following coupled differential equations can be obtained for diagonal matrix elements,  
\begin{eqnarray}
&&\dot{\rho_{11}}= \Gamma_{abs} (\rho_{22} + \rho_{33}) + \Gamma_{dph}^x(-2\rho_{11} + \rho_{22} + \rho_{33}) - \frac{J}{2}\rho_{14},\nonumber \\
&&\dot{\rho_{22}}= \Gamma_{abs} (-\rho_{22} + \rho_{44}) + \Gamma_{dph}^x(\rho_{11} - 2\rho_{22} + \rho_{44}) - \frac{J}{2}\rho_{23},\nonumber \\
&&\dot{\rho_{33}}= \Gamma_{abs} (-\rho_{33} + \rho_{44})  + \Gamma_{dph}^x(\rho_{11} - 2\rho_{33} + \rho_{44})+ \frac{J}{2}\rho_{23},\nonumber\\
&&\dot{\rho_{44}}= -2 \Gamma_{abs} \rho_{44} + \Gamma_{dph}^x(\rho_{22} + \rho_{33} -2\rho_{44}) + \frac{J}{2}\rho_{14}. \nonumber \\
\end{eqnarray}
and for off-diagonal ones,
\begin{eqnarray}
&&\dot{\rho_{12}}= \Gamma_{abs} (\rho_{34} - \frac{\rho_{12}}{2}) + \Gamma_{dph}^x (\rho_{34}-\rho_{12}) + \mbox{Im},\nonumber\\
&&\dot{\rho_{13}}= \Gamma_{abs} (\rho_{24} - \frac{\rho_{13}}{2}) + \Gamma_{dph}^x (\rho_{24}-\rho_{13}) + \mbox{Im} \nonumber\\
&&\dot{\rho_{14}}= -\Gamma_{abs} \rho_{14} + 2\Gamma_{dph}^x (-\rho_{14}+ \rho_{23}) +  \mbox{Im},\nonumber\\
&&\dot{\rho_{23}}= -\Gamma_{abs} \rho_{23} + 2\Gamma_{dph}^x(-\rho_{23}+ \rho_{14}) + \mbox{Im},\nonumber \\
&&\dot{\rho_{24}}= (- 3\Gamma_{abs}/2 - \Gamma_{dph}^x) \rho_{24} + \Gamma_{deph} \rho_{13} +  \mbox{Im}, \nonumber\\
&& \dot{\rho_{34}}= (-3\Gamma_{abs}/2 -  \Gamma_{dph}^x) \rho_{34} + + \Gamma_{deph} \rho_{12}+  \mbox{Im}. \nonumber\\
\end{eqnarray}
The notations are similar to those used in the phase-flip noise. Again similar simplification leads to \(W(\Gamma_{dph}^x)\).


\begin{thebibliography}{100}

\bibitem{NielsenChuangbook} M. Nielsen and I. Chuang, \emph{Quantum Computation and Quantum Information} (Cambridge University Press,
Cambridge, 2000).





\bibitem{pleniorev} T. Baumgratz, M. Cramer, and M. B. Plenio, Phys. Rev. Lett. {\bf 113}, 140401 (2014); A. Streltsov, G. Adesso, and M. B. Plenio,
Rev. Mod. Phys. {\bf 89}, 041003  (2017). 

\bibitem{horodecki} R. Horodecki, P. Horodecki, M. Horodecki, and K. Horodecki, Rev. Mod. Phys. {\bf 81}, 865 (2009).

\bibitem{amader_ent} S. Das, T. Chanda, M. Lewenstein, A. Sanpera, A. Sen(De), and U. Sen, \emph{The separability versus entanglement problem}, in \emph{Quantum Information: From Foundations to Quantum Technology Applications}, second edition, eds. D. Bruß and G. Leuchs (Wiley, Weinheim, 2019), arXiv:1701.02187 [quant-ph].

\bibitem{modiamader} K. Modi,  A. Brodutch, H. Cable, T. Paterek, and V. Vedral, Rev. Mod. Phys. {\bf 84}, 1655 (2012); A. Bera, T. Das, D. Sadhukhan, S. Singha Roy, A. Sen(De) and U. Sen, Rep. Prog. Phys. {\bf 81}, 024001 (2018).

\bibitem{revcrypto} N. Gisin, G. Ribordy, W. Tittel, and H. Zbinden, Rev. Mod. Phys. {\bf 74}, 145 (2002); V. Scarani, H. Bechmann-Pasquinucci, N. J. Cerf, M. Dusek, N. Lutkenhaus, and M. Peev, Rev. Mod. Phys. {\bf 81}, 1301 (2009) ; M. Krenn, M. Malik, T. Scheidl, R. Ursin, and A. Zeilinger, Optics in Our Time (pp. 455-482), Springer International Publishing (2016). 
%U. L. Andersen, L. Banchi, M. Berta, D. Bunandar, R. Colbeck, D. Englund, T. Gehring, C. Lupo, C. Ottaviani, J. Pereira, M. Razavi, J. S. Shaari, M. Tomamichel, V. C. Usenko, G. Vallone, P. Villoresi, P. Wallden

\bibitem{oneway} R. Raussendorf and H. J. Briegel, Phys. Rev. Lett. {\bf 86}, 5188 (2001). 


\bibitem{optcontrol} D.  Dong and I. R Petersen, IET Control Theory and Applications, {\bf 4}, 2651 (2010).

\bibitem{photonRMP} J.-W. Pan, Z.-B. Chen, C.-Y. Lu, H. Weinfurter, A. Zeilinger, and M. {\.Z}ukowski, 
Rev. Mod. Phys. {\bf 84}, 777 (2012). 

\bibitem{ionRMP} L.-M. Duan and C. Monroe,
Rev. Mod. Phys. {\bf 82}, 1209 (2010). 

\bibitem{alicki} R. Alicki and M. Fannes, Phys. Rev. E {\bf 87}, 042123 (2013).
\bibitem{campaioli'18} F. Campaioli, F. A. Pollock, and S. Vinjanampathy, arXiv:1805.05507 [quant-ph] (2018).

\bibitem{qthermobook} J. Gemmer, M. Michel, and G. Mahler, \emph{Quantum Thermodynamics: Emergence of Thermodynamic Behavior Within Composite Quantum Systems, Lect. Notes Phys. 784} (Springer, Berlin Heidelber 2009). 

\bibitem{horojona} M. Horodecki, and J. Oppenheim, Nat. Comm. {\bf  4},  2059 (2013).

\bibitem{acin} K. V. Hovhannisyan, M. Perarnau-Llobet, M. Huber, and A. Acin, Phys. Rev. Lett. {\bf 111}, 240401 (2013).

\bibitem{binder} F. Binder, S. Vinjanampathy, K. Modi, and J. Goold, New J. Phys. {\bf 17}, 075015 (2015).

\bibitem{andolina1} G.M. Andolina, M. Keck, A. Mari, M. Campisi, V. Giovannetti, and M.  Polini,
Phys. Rev. Lett. {\bf 122}, 047702 (2019).


 

\bibitem{modisai'17} F. Campaioli, F. A. Pollock, F. C. Binder, L. Celeri, J. Goold, S. Vinjanampathy, and K. Modi, Phys. Rev. Lett. {\bf 118}, 150601 (2017).

\bibitem{Qvscl}G.M. Andolina, M. Keck, A. Mari, V. Giovannetti, and M. Polini,
Phys. Rev. B {\bf 99}, 205437  (2019). 


\bibitem{ferraro} D. Ferraro, M. Campisi, G. M. Andolina, V. Pellegrini, and M. Polini, Phys. Rev. Lett. {\bf 120}, 117702 (2018).

\bibitem{lemodi'18} T. P. Le, J. Levinsen, K. Modi, M. M. Parish, and F. A. Pollock, Phys. Rev. A {\bf 97}, 022106 (2018).

\bibitem{andolina'18} G. M. Andolina, D. Farina, A. Mari, V. Pellegrini, V. Giovannetti, and M. Polini, Phys. Rev. B {\bf 98}, 205423 (2018).


\bibitem{rosa19} D. Rossini, G.M.  Andolina, D. Rosa, M.  Carrega, M.  Polini, Phys. Rev. Lett. {\bf 125}, 236402 ; D.  Rosa, D. Rossini, G.M.  Andolina, M. Polini, and M. Carrega,Journal of High Energy Physics volume 2020, {\bf 67} (2020).

\bibitem{Crescente} A. Crescente, M. Carrega, M. Sassetti, D. Ferraro,
New J. Phys. {\bf 22} 063057 (2020).

\bibitem{srijon'20} S. Ghosh, T. Chanda, and A. Sen(De), Phys. Rev. A {\bf 101}, 032115 (2020).

\bibitem{rossini'19} D. Rossini, G.M. Andolina, and M. Polini, Phys. Rev. B { \bf 100}, 115142  (2019).  



 \bibitem{opensysbook} H. P. Breuer and F. Petruccione, \emph{The Theory of Open Quantum Systems} (Oxford University Press, Oxford, 2007).


\bibitem{farina'19} D. Farina, G. M. Andolina, A. Mari, M. Polini, and V. Giovannetti, Phys. Rev. B {\bf 99}, 035421 (2019).

\bibitem{alickiopen}  R. Alicki, \emph{A quantum open system model of molecular battery charged by excitons}, J. Chem. Phys. {\bf 150}, 214110 (2019).

\bibitem{latune'19} C. L. Latune, I. Sinayskiy, and F. Petruccione, Phys. Rev. A {\bf 99}, 052105 (2019).

\bibitem{kamin'19}  F. H. Kamin, F. T. Tabesh, S. Salimi, F. Kheirandish, and A.  C. Santos, New J. Phys. {\bf 22} (2020) 083007.

\bibitem{zakavati'20} S. Zakavati, F. T. Tabesh, and S. Salimi, arXiv: arXiv:2003.09814. 

\bibitem{Gherardini'20} S. Gherardini, F. Campaioli, F. Caruso, and F. C. Binder,  Phys. Rev. Research {\bf 2}, 013095  (2020).


\bibitem{munro'20} J. Q. Quach, and W. J. Munro, Phys. Rev. Applied {\bf 14}, 024092 (2020).

\bibitem{sibai_2020} Si-Yuan Bai and Jun-Hong An,
Phys. Rev. A {\bf 102}, 060201(R).

\bibitem{mitchison} Mark T. Mitchison, John Goold, and Javier Prior,
Qunatum {\bf 5} 500.

\bibitem{carrega}  M Carrega, A Crescente, D Ferraro and M Sassetti,
New J. Phys. {\bf 22} 083085. 

%\bibitem{kamin'20} F. H. Kamin, F. T. Tabesh, S. Salimi, F. Kheirandish, and A.  C. Santos, arXiv:1910.07751 (2020).

\bibitem{breuer'09} H.-P. Breuer, E.-M. Laine, and J. Piilo, Phys. Rev. Lett. {\bf 103}, 210401 (2009).

\bibitem{thorwart'09} M. Thorwart, J. Eckel, J. H. Reina, P. Nalbach, and S. Weiss, Chem. Phys. Lett. {\bf 478}, 234 (2009).

\bibitem{laine'10}  E.-M. Laine, J. Piilo, and H.-P. Breuer, Phys. Rev. A {\bf 81}, 062115 (2010).

\bibitem{rivas'10} A. Rivas, S. F. Huelga, and M. B. Plenio, Phys. Rev. Lett. {\bf 105}, 050403 (2010).

\bibitem{lu'10}  X.-M. Lu, X. Wang, and C. P. Sun, Phys. Rev. A { \bf 82}, 042103 (2010).


\bibitem{vasile'11} R. Vasile, S. Olivares, M. G. A. Paris, and S. Maniscalco, Phys. Rev. A {\bf 83}, 042321 (2011).





\bibitem{schmidt'11} R. Schmidt, A. Negretti, J. Ankerhold, T. Calarco, and J. T. Stockburger, Phys. Rev. Lett. {\bf 107}, 130404 (2011).

\bibitem{huelga'12} S. F. Huelga, A. Rivas, and M. B. Plenio, Phys. Rev. Lett. {\bf 108}, 160402 (2012).

\bibitem{Chin'12} A. W. Chin, S. F. Huelga, and M. B. Plenio, Phys. Rev. Lett. {\bf 109}, 233601 (2012).

\bibitem{haikka'13} P. Haikka, T. H. Johnson, and S. Maniscalco, Phys. Rev. A {\bf 87}, 010103(R) (2013). 

\bibitem{sabrinarev} G. Karpat, C. Addis, and S. Maniscalco, \emph{Lectures on General Quantum Correlations and their Applications. Quantum Science and Technology} (Springer, 2017) pp 339-366.

\bibitem{titassamya} T. Chanda, and  S. Bhattacharya,  Ann.  Phys.  {\bf 366}, 1 (2016). 

\bibitem{rivu'20} R. Gupta, S. Gupta, S. Mal, and A. Sen (De), arXiv: 2005.04009.

\bibitem{Sachdevbook} S. Sachdev, \emph{Quantum Phase Transitions} (Cambridge University Press, Cambridge, 2011).


\bibitem{lindblad'76} G. Lindblad, Commun. Math. Phys. {\bf 48}, 119 (1976).


 
 \bibitem{gorini'76} V. Gorini, A. Kossakowski, and E. C. G. Sudarshan, J. Math. Phys. {\bf 17}, 821 (1976).




\bibitem{vidal'02} G. Vidal, and R.F. Werner, Phys. Rev. A {\bf 65}, 032314 (2002).

\bibitem{LN} Logarithmic negativity \cite{vidal'02} of a two-qubit density matric, \(\rho_{AB}\) can be obtained by the absolute value of negative eigenvalue of the partial transposed state, \(\rho_{AB}^{T_A}\) with partial transposition being taken with respect to \(A\) \cite{PT}. 

\bibitem{PT} A. Peres, Phys. Rev. Lett. {\bf 77}, 1413 (1996);  M. Horodecki, P. Horodecki and R. Horodecki, Phys. Lett. A {\bf 223}, 1 (1996).


\bibitem{fazioreview} L. Amico, R. Fazio, A. Osterloh, and V. Vedral, Rev. Mod. Phys. {\bf 80},  517  (2008).

\bibitem{aditireview} M. Lewenstein, A. Sanpera, V. Ahufinger, B. Damski,
A. Sen(De), and U. Sen, Adv. Phys. {\bf 56}, 243 (2007); M. Lewenstein, A. Sanpera, and V. Ahufinger, \emph{Ultracold atoms in Optical Lattices: simulating quantum many-body physics} (Oxford University Press, Oxford, 2012).
%**** open QB*****


%nonMarko realizable




\bibitem{worknew} P. Haikka, S. McEndoo, G. De Chiara, G. M. Palma, and
S. Maniscalco, Phys. Rev. A {\bf 84}, 031602 (2011); F. Cosco, M. Borrelli, J. J. Mendoza-Arenas, F.Plastina, D. Jaksch, and S. Maniscalco, Phys. Rev. A {\bf 97}, 040101(R) (2018); A. Lampo, C. Charalambous, M. {\'A}. Garc\'{\i}a-March, and M. Lewenstein, Phys. Rev. A {\bf 98}, 063630  (2018) and references therein. 

\bibitem{titas} \url{https://titaschanda.github.io/QIClib}. 



\end{thebibliography}
\end{document}